\documentclass[fleqn,usenatbib]{mnras}

\usepackage{newtxtext,newtxmath}

\usepackage[T1]{fontenc}
\usepackage{ae,aecompl}

\usepackage{graphicx}	
\usepackage{amsmath}	
\usepackage{amssymb}	
\usepackage{longtable}    

\newcommand{\Cratio}{$^{12}$C$/$$^{13}$C\,}	

\title[Carbon Fractionation in Galaxies]{A chemical study of carbon fractionation in external galaxies}

\author[Serena Viti et al.]{
Serena Viti$^{1,2}$\thanks{E-mail: viti@strw.leidenuniv.nl},
Francesco Fontani$^{3,4}$,
Izaskun Jim\'enez-Serra$^{5,2}$,
\\
$^{1}$ Leiden Observatory, Leiden University, PO Box 9513, 2300 RA Leiden, The Netherlands \\
$^{2}$Department of Physics and Astronomy, University College London,
Gower Street, London, WC1E 6BT, UK\\
$^{3}$Osservatorio Astrofisico di Arcetri, Largo E. Fermi 2, 
I-50125 Firenze, ITALY\\
$^{4}$ Centre for Astrochemical Studies, Max-Planck-Institute for Extraterrestrial Physics, Giessenbachstrasse 1, 85748 Garching, Germany \\
$^{5}$Centro de Astrobiolog\'ia (CSIC/INTA),
Ctra de Torrej\'on a Ajalvir, km 4
28850 Torrej\'on de Ardoz, Madrid Spain}

\date{Accepted 2020 July 21. Received 2020 July 14; in original form 2019 October 31}

\pubyear{2019}

\begin{document}
\label{firstpage}
\pagerange{\pageref{firstpage}--\pageref{lastpage}}
\maketitle

\begin{abstract}
In the interstellar medium carbon exists in the form of two stable isotopes $^{12}$C and $^{13}$C and their ratio is a good indicator of nucleosynthesis in galaxies. However, chemical fractionation can potentially significantly alter this ratio and in fact observations of carbon fractionation within the same galaxy has been found to vary from species to species. In this paper we theoretically investigate the carbon fractionation for selected abundant carbon-bearing species in order to determine the conditions that may lead to a spread of the \Cratio  ratio in external galaxies. We find that carbon fractionation is  sensitive to almost all the physical conditions we investigated, it strongly varies with time for all species but CO, and shows pronounced differences across species. Finally we discuss our theoretical results  in the context of the few observations of the \Cratio in both local and higher redshift galaxies.

\end{abstract}

\begin{keywords}
Galaxies: ISM -- ISM: abundances -- ISM: molecules --ISM: fractionation
\end{keywords}



\section{Introduction}

Knowledge of isotopic abundances is important in galaxy evolution studies because isotopes provide diagnostics for the chemical enrichment in galaxies over time. While ideally measurements of isotopes in large sample of stars are needed to determine the fossil record of the enrichment history, in practice this is hampered by the need of very high resolution, high signal-to-noise spectroscopic data. A complementary, or alternative, method is to measure isotopic ratios from observations of gas-phase interstellar medium (ISM) isotopic abundances (Wilson \& Rood 1994).

Carbon is one of the most abundant and important elements in the Universe, and exists in the form of two stable isotopes, $^{12}$C 
and $^{13}$C. Its isotopic composition is believed to be a good indicator of nucleosynthesis in Galaxies. $^{12}$C is predicted to be
synthesised rapidly via Helium burning in both low-mass and massive stars, and it is hence considered a primary element, even though the relative contributions from low- and intermediate-mass stars and from massive stars change with time and position within the Galaxy (e.g.~Akerman et al. 2004, Romano et al.~2017). $^{13}$C is mainly expected
to be synthesised through the CNO cycle in asymptotic giant branch (AGB) stars through slower processes, and it is a secondary 
element (e.g.~Wilson \& Rood 1994, Pagel~1997). After $^{13}$C is created in the interior of these stars, processes like convection
instabilities and expansion of the outer layers can pull $^{13}$C to the stellar surface and there, mixed in the stellar atmosphere, 
could be ejected into the surrounding ISM.  However, other processes unrelated to stellar nucleosynthesis, such as
chemical fractionation, could potentially  significantly affect the isotopic composition of the ISM. 
Various chemical mechanisms that may introduce deviations from
the carbon isotopic elemental ratio have been investigated.
Several works proposed low-temperature ion-neutral exchange
reactions (Langer et al. 1984, Smith \& Adams 1980, Woods
\& Willacy 2009, Mladenovic \& Roueff 2014) which, owing to their exothermicity, can increase $^{13}$C abundance in CO and HCO$^+$ in very
low ($T\simeq 10$ K) temperature environments, similarly to the chemical
mechanism invoked to explain the high abundances of deuterated
species in cold gas (Watson et al. 1976). However, this simple
picture has been recently revised because not all these reactions
occur even at very low temperatures, due to the existence of energy barriers between the interacting particles (Roueff et al. 2015).

Alternative mechanisms have been explored in regions different
from the cold and dense environments required for fractionation
dominated by low-temperature ion-neutral reactions. In
photo-dissociation regions (PDRs), or diffuse regions exposed
to UV radiation fields (e.g.~Liszt 2007, Visser et al. 2009,
R\"{o}llig \& Ossenkopf 2013), isotope-selective photo-dissociation
and other mechanisms are claimed to influence, and potentially
change significantly, the carbon elemental isotopic ratios.
These mechanisms can dominate over others in galaxies characterised
by strong radiation fields.

Effects of fractionation have been evaluated
in molecular clouds of the Milky Way by measuring \Cratio\ from several molecular species such as CN, CO, HCN and H$_2$CO and the Galactic gradient in \Cratio\ is found to depend on the species used (e.g.~Wilson \& Rood 1994, Wilson~1999, Savage et al.~2002, Milam et al.~2005), which indeed suggests additional fractionation effects on top of local isotopic abundances due to Galactic nucleosynthesis. 

Recently, Roeuff et al. (2015) developed a chemical model which included a comprehensive $^{13}$C network and simulated the gas evolution of standard galactic molecular clouds. They found large variations in the $^{12}$C/$^{13}$C ratio across time as well as across molecules, a consequence of the many different reaction channels involved in the $^{13}$C chemistry. Therefore, it is likely that, as we move away from standard galactic conditions, such ratio will display even larger variations among molecules and type of galaxies. 

There have been several  observational studies of the \Cratio in external galaxies (e.g. Henkel et al. 1998, 2010, 2014; Mart\'in et al. 2005, 2006, 2019; Gonz\'alez-Alfonso et al. 2012; Aladro et al. 2013; Wang et al. 2009; Jiang et al. 2011; Tang et al. 2019; see also Table 2). The molecules most used to determine this ratio in external galaxies are: CN (found to be the best tracer by Henkel et al. 2014), CO, H$_2$CO, HCO$^+$, HCN and CS. Such observations confirm the large variations found in our own Galaxy: in fact within active galaxies (starburst or ULIRG) the observed range encompasses a full order of magnitude (Henkel et al. 2014), although the observed sample  is rather small.  While the observations have concentrated on galaxies where the energetics are different from those of our own Galaxy,  a theoretical study exploring the dependence of carbon chemical fractionation on such energetics and various physical characteristics
is, so far, missing. 

In Viti et al.~(2019) [hereafter Paper I], we have investigated the effect of nitrogen fractionation in external galaxies to determine the physical
conditions that may lead to significant changes in the $^{14}$N/$^{15}$N ratio from reference values (e.g.~Solar, or Galactic
values, both around 400), and found that the main cause of ISM increase of nitrogen fractionation is high gas densities, strengthened   
by high fluxes of cosmic rays.
In this study, we perform, for the first time, a chemical modelling study of carbon fractionation that 
may be occurring in the gaseous component of external galaxies by updating the model presented in Paper I. 
We evaluate the effect of carbon fractionation for selected abundant C-bearing 
species, and also investigate if having neglected these reactions has had significant effects on the nitrogen fractionation. In section 2, we present the 
chemical model and network used for the $^{12}$C and $^{13}$C isotopic species; in Section 3 we present our results for the modelling of the carbon 
fractionation in gas at different H$_2$ densities and extinction, and affected by energetic phenomena (such as stellar heating, UV radiation and cosmic rays). 
In Section 4, we briefly report our conclusions.

\section{Chemical modelling of carbon fractionation}
\label{model}

As in Paper I, we used the open source time dependent gas-grain chemical code UCLCHEM\footnote{https://uclchem.github.io/}.   UCLCHEM computes the time evolution of the chemical abundances of the gas and of the species on the ices.   We note that since the study published in Paper I, updates were made to the UCLCHEM code, including a better treatment of the charge conservation. This only affects models with very high cosmic ray ionization rates, where in any case the chemistry is better modelled by codes where the thermal balance is calculated (see later). We  have in fact compared the abundances of the main nitrogen-bearing species, their isotopologues and the fractionation ratios and found that in small cases the behavioural trends remain unchanged. 

We ran UCLCHEM in two phases in a very similar manner as in Viti (2017) where theoretical abundances for extragalactic studies were derived.  In Phase I, the gas is allowed to collapse from a diffuse atomic state ($\sim$ 10 cm$^{-3}$) up to a high density (chosen by the user) by means of a free-fall collapse. The temperature during Phase I is always 10 K, and the cosmic ray ionization rate and radiation field are at their standard Galactic values of $\zeta_o$ = 5$\times$10$^{-17}$ s$^{-1}$ and 1 Draine, (equivalent to 1.6$\times$10$^{-3}$ erg/s/cm$^2$; Draine~1978, Draine \& Bertoldi~1996), respectively. During Phase I, atoms and molecules  freeze onto the dust  and react with each other, forming icy mantles.  For this grid of models we choose to end Phase I when the final density is reached.  Phase II  then follows  the chemical evolution of this gas for 10$^7$ years after some energetic event has occurred (simulating either the presence of an AGN and/or a starburst). 
The initial (solar) elemental abundances considered in our models were taken from Asplund et al. (2009). Our elemental isotopic carbon ratio is 69 (Wilson 1999). Further details of the UCLCHEM code can be found in Holdship et al. (2017) and also in Paper I.


In both phases, the basic gas phase chemical network is the UMIST13 database (McElroy et al. 2013), augmented and updated with the KIDA database (Wakelam et al. 2015). The surface reactions included in the modelling for this study are mainly hydrogenation reactions, 
allowing chemical saturation when possible. The network contains 4483 reactions and 338 chemical species. \\

\begin{figure}
    \centering
    \includegraphics[scale=0.50]{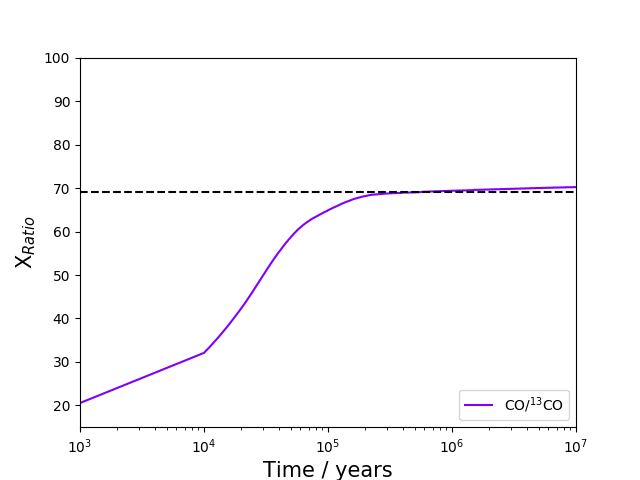}
    \includegraphics[scale=0.50]{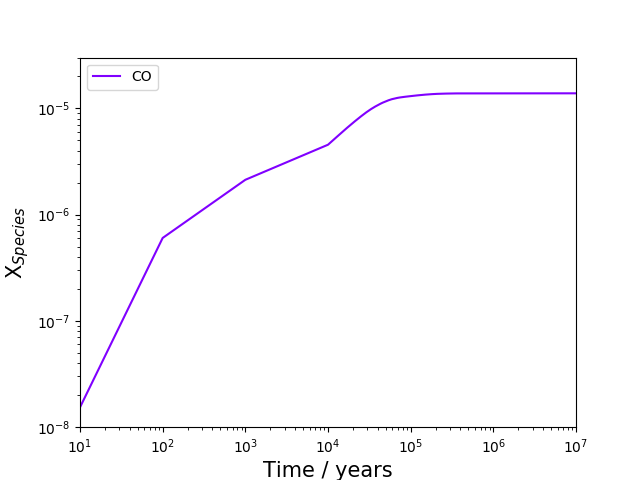}
    \caption{The $^{12}$CO/$^{13}$CO ratio (Top) and the CO fractional abundance (Bottom) for our benchmarking model (see text). }
    \label{fig:test}
\end{figure}

For the $^{13}$C network, we duplicated the $^{12}$C network changing all $^{12}$C by $^{13}$C,  as we assume that reactions involving isotopic molecules have the same total rate constant as those involving the main isotope (Roeuff et al. 2015). We also added the $^{13}$C exchange reactions reported by Roueff et al. (2015) in their Table 1. However, for this initial study and to limit the size of the network, we  only considered the fractionation of one carbon atom in a molecule where multiple C atoms are present. This of course may lead to inaccuracies but only probably in minor species, not routinely observed in extragalactic environments. This is confirmed by our benchmarking test with models from Roeuff et al. (2015): in order to estimate the potential shortcomings of our reduced network we ran a model using the same initial conditions of their Model (b) [see Table 4 in Roeuff et al. (2015)]. Note  that while we adopted the same elemental abundances as well as the same physical conditions as in their model, our network is different (in terms of species as well as reaction network) and some of our initial conditions may differ (e.g. the initial ionization fraction). Hence we do not expect a perfect match in absolute abundances. However, to illustrate the overall consistency between the two models, in Figure~\ref{fig:test} we show the \Cratio ratio for CO  (to be compared with their Figure 5 top right)
and the CO abundance (to be compared with their Figure 5 bottom left, although note that Roueff et al. (2015) abundances are with respect to molecular hydrogen and $not$ the total number of hydrogen nuclei as they are here). It can be seen that, although our model has a higher CO abundance and \Cratio at earlier times, both our equilibrium \Cratio ratio and the CO abundance are in agreement (our equilibrium CO fractional abundance with respect to the total number of hydrogen nuclei is 1.38$\times$10$^{-5}$, and Roueff et al. (2015)'s equilibrium abundance is 1.39$\times$10$^{-5}$). As the scope of this paper is to study qualitative trends in large scale gas in extragalactic environment, we do not at present perform any more detailed benchmarking with previous models.  
\par 
The grid of models we ran is very similar, in terms of parameter space, to the one ran in Paper I, namely we varied: i) gas densities (n$_H$) from 10$^4$ to 10$^6$ cm$^{-3}$; ii) visual extinctions (A$_V$) from 1 to 100 mags; iii) temperatures (T) of 50 and 100 K; iv) radiation fields ($\chi$) from 1 to 1000 Draine; v) cosmic ray ionization rates ($\zeta$) from 1 to 10$^4$ the standard galactic one. As explained in detail in Paper I such parameter space is supposed to qualitatively represent the range of conditions found in external galaxies,  whereby for example high radiation fields, visual extinctions and temperatures are representative of starburst galaxies, while high cosmic ray ionization rates are typically found in AGN-dominated galaxies. However, we in fact note that, at cosmic rays ionization rates as high as 10,000 times the standard one, the temperature of the gas may be  higher than 100 K and hence some of the chemistry may be  partially suppressed. The driver of this decline in molecular abundances is linked to the decline in molecular hydrogen  as discussed in Bayet et al. (2011). In order to briefly test this we ran two test cases using an updated preliminary version of UCLCHEM where thermal balance is included (Holdship et al. in preparation) and found that  at densities of 10$^5$ cm$^{-3}$, visual extinctions $>$ 1 mag, and cosmic ray ionization rates 1000 and 10000 times the standard one, the temperatures are respectively 70 and 140 K (J. Holdship, private communication), both close enough to our adopted temperatures to  not change the qualitative trends of our results.
\section{Results}
\label{results}
The main aim of our study is to understand how carbon fractionation varies across a parameter space covering physical characteristics and energetics representing different types of galaxies for different species. We report these findings in Sections 3.1 and 3.2. While we do not attempt at modelling any particular galaxy, we compare our theoretical C fractionation with that found in a handful of galaxies in Section 3.3. 

\subsection{Dependence of carbon fractionation to variations of the physics and energetics of the galaxies}
\label{dependence}
Many theoretical studies have investigated the sensitivity of chemistry to the range of physical parameters that characterize a galaxy (e.g Bayet et al. 2008, 2009, 2011; Meijerink et al. 2011; Kazandjian et al. 2012, 2015). In this paper we only concentrate on discussing these effects on the $^{12}$C/$^{13}$C ratio for various species.  Nevertheless, for the species and models for which this ratio is discussed, we also plot the fractional abundances of the main isotopologues in order to show that  they are indeed detectable  (see Figure~\ref{fig:abundances}). Note that all the plots are for Phase II of the models. We limit the species to the molecules most likely to be observed in extragalactic conditions. In addition, we also plot the $^{12}$C/$^{13}$C ratio for C and C$^+$ as they are often the drivers of changes in the fractionation of the molecules we discuss.  

\begin{figure*}
    \centering
    \includegraphics[scale=0.50]{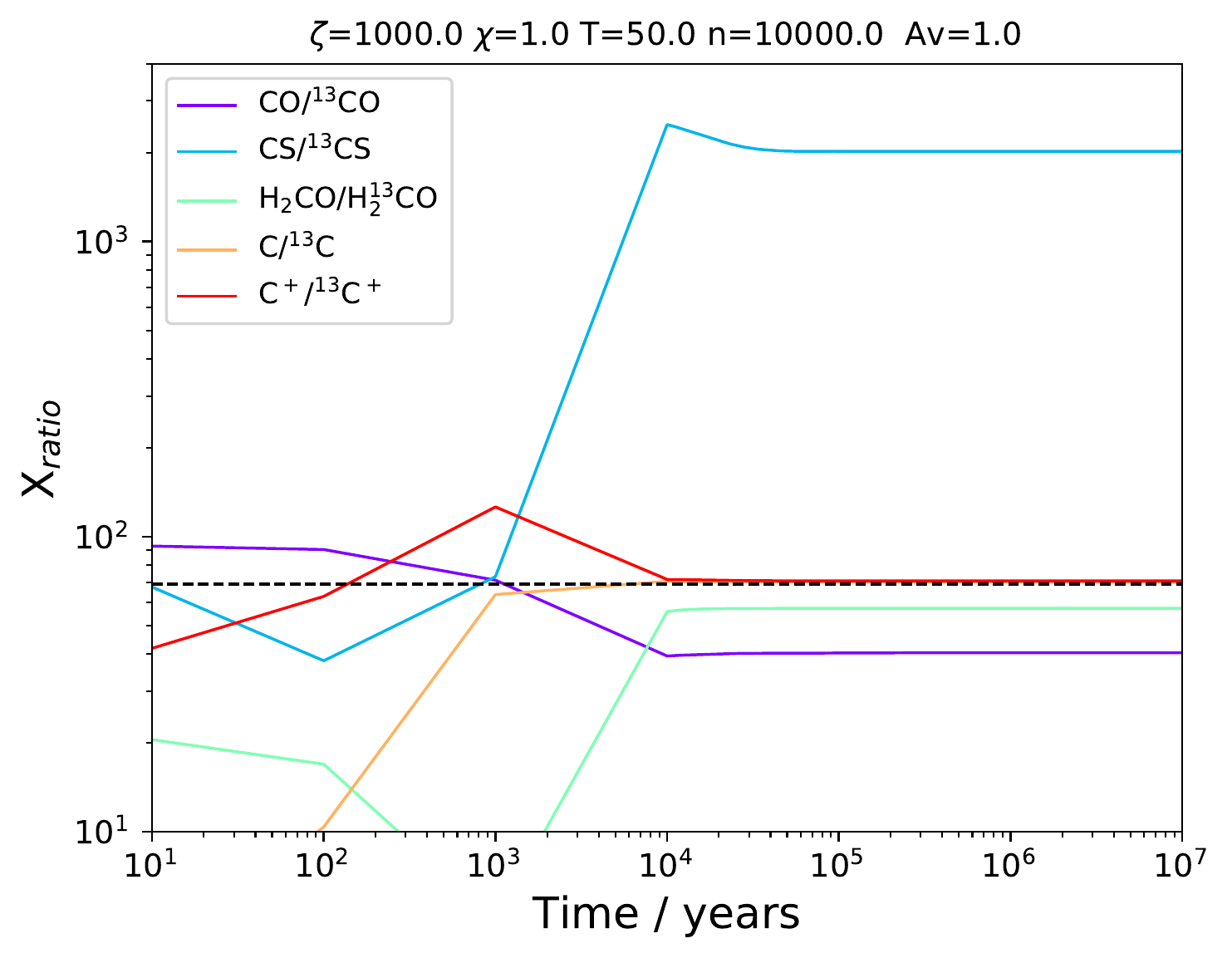}
    \includegraphics[scale=0.50]{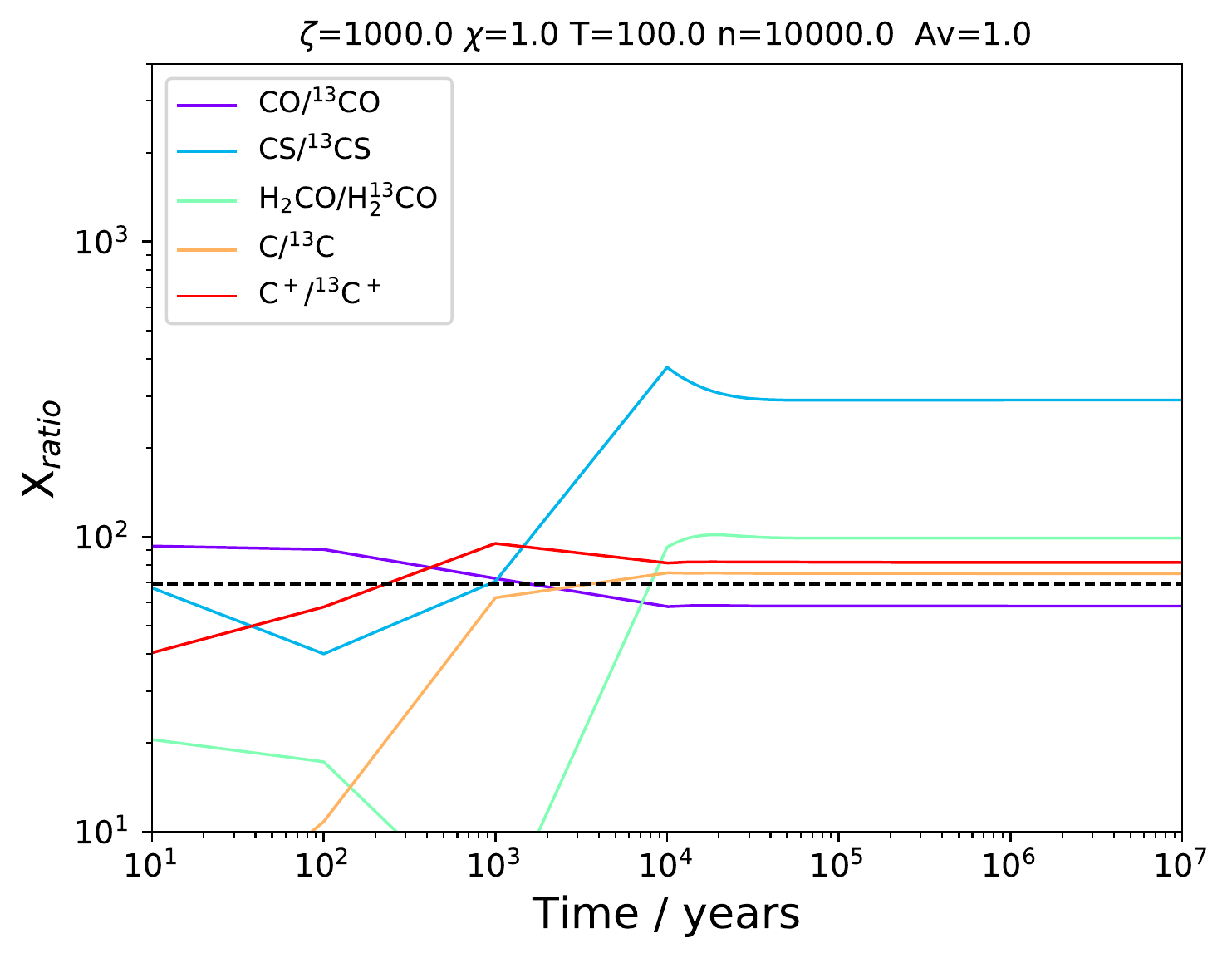}
    \includegraphics[scale=0.50]{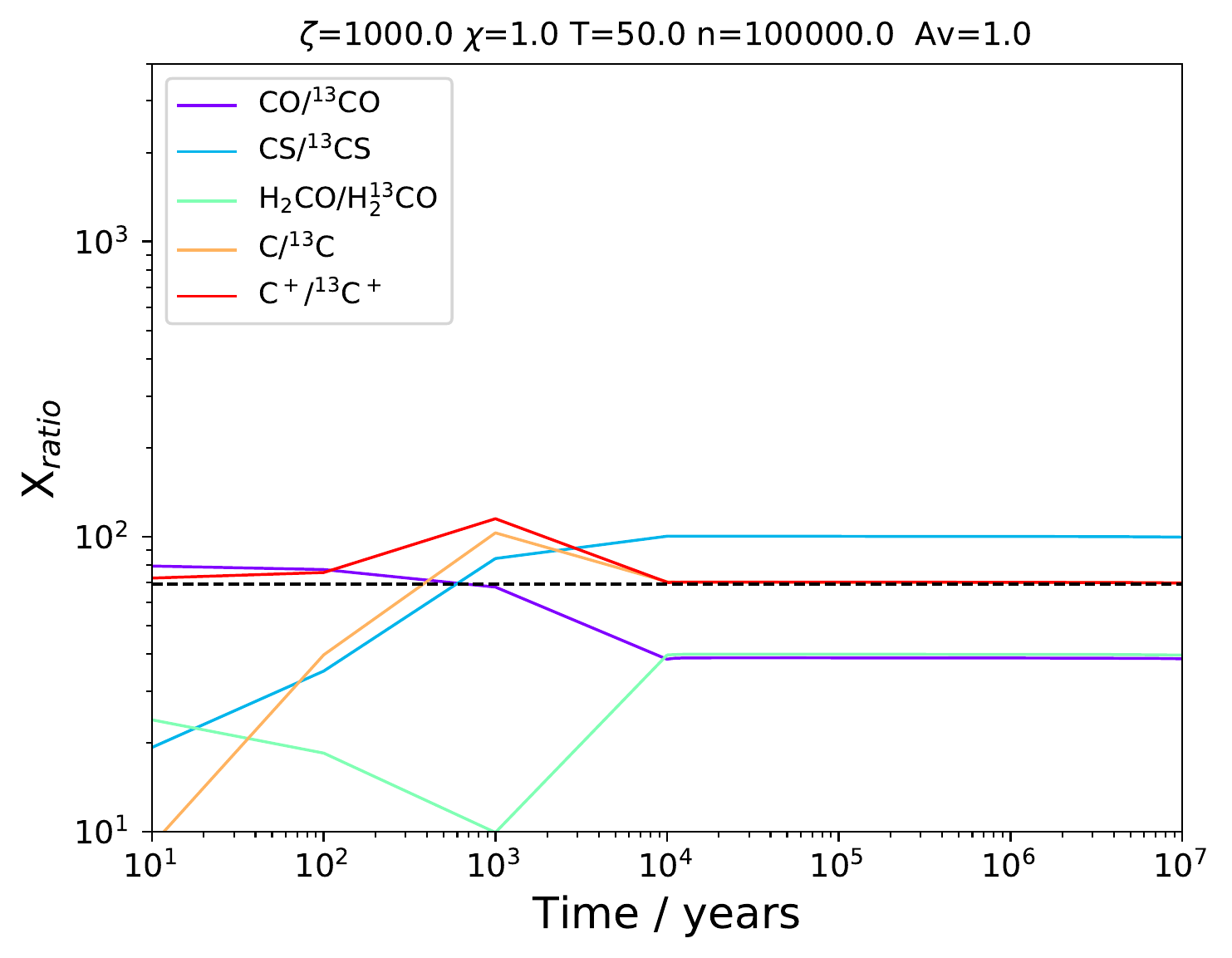}
    \includegraphics[scale=0.50]{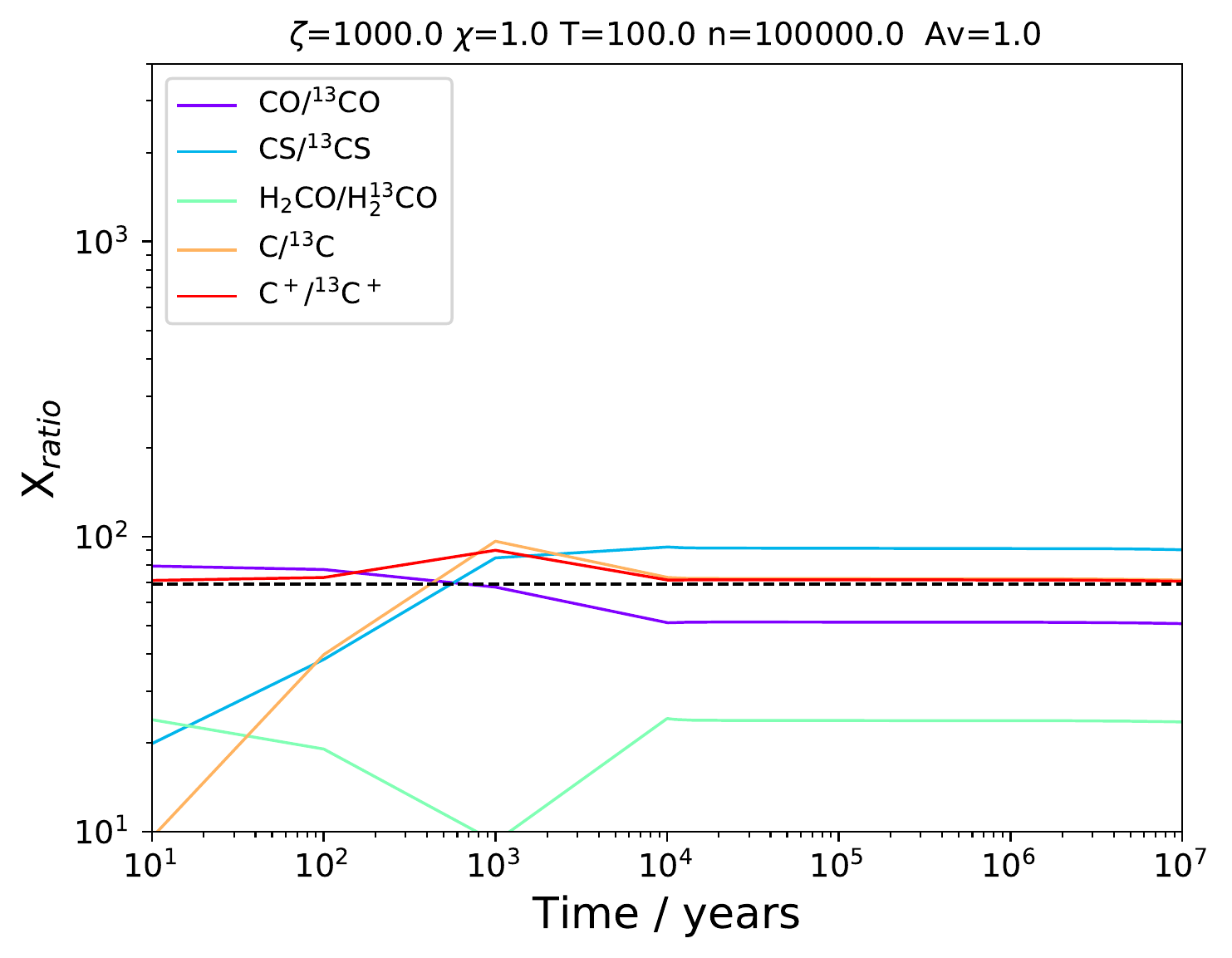}
    \caption{Example of differences in C fractionation due to the increase in gas temperature. In this Figure, as in all the others, the horizontal dashed line indicates the reference value of 69 for the \Cratio in the local ISM (Wilson~1999). The units of the parameters in the title for each plot of this Figure and of all the following Figures are: $\zeta_o$, Draine, K, cm$^{-3}$ and magnitudes. }
    \label{fig:Tvar}
\end{figure*}

As in Paper I, we ran all our models at two different temperatures, 50 and 100~K. We find, just as for the nitrogen fractionation, that temperature differences do not affect or affect only marginally carbon fractionation (depending on the molecule, in the most 'extreme' cases, by a factor of 2 or so at most). At low cosmic rays ionization rates,
there are no significant differences. There are some small differences between the two sets of models, occurring  at late times ($\geq$ 10$^6$ years), when the cosmic rays ionization rate is high ($\zeta \geq 1000$), the gas density is low (10$^4$ or 10$^5$ cm$^{-3}$) and the visual extinction is $\sim$ 1 mags, namely mainly a decrease in the \Cratio  at higher temperatures. Examples of the most `extreme' variations due to temperatures are plotted in Figure~\ref{fig:Tvar}.  

Beside that we note the following:  
\begin{itemize}
\item the $^{12}$CO/$^{13}$CO  ratio is surprisingly constant with time and across the parameter space we investigated. In terms of actual value, it rarely reaches a minimum of 30 or a maximum of 100, usually remaining constant at $\sim$ 70. The lowest values  are only ever reached for models at low density, with A$_V$ = 1 mags, and a high radiation field (or a combination of high radiation fields and high cosmic ray ionization rates). Higher values for $^{13}$CO are very rare and only for very early times ($<$ 1000 yrs), low densities and low visual extinctions (see Fig~\ref{fig:CO}).  

\begin{figure}
    \centering
    \includegraphics[scale=0.60]{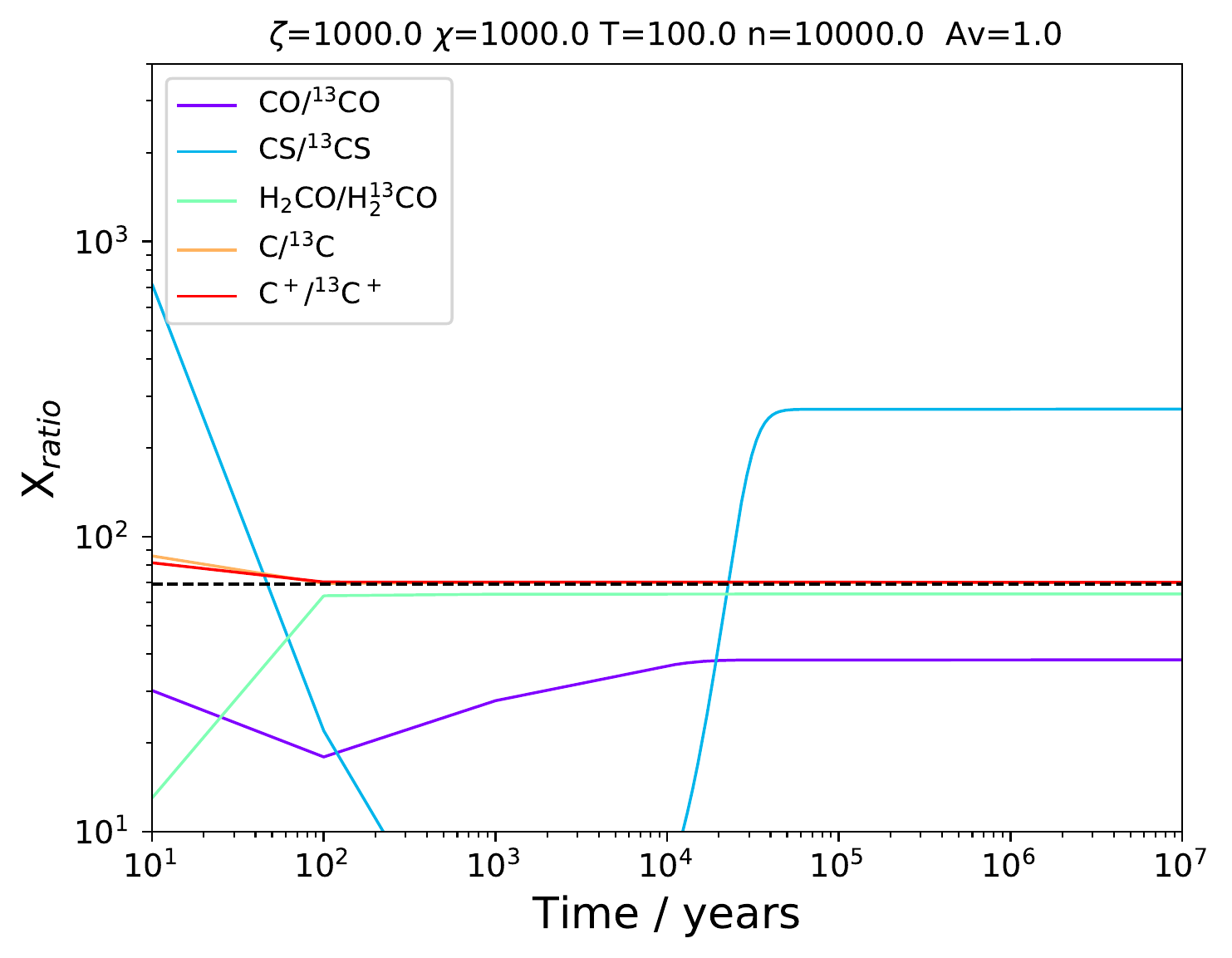}
    \includegraphics[scale=0.60]{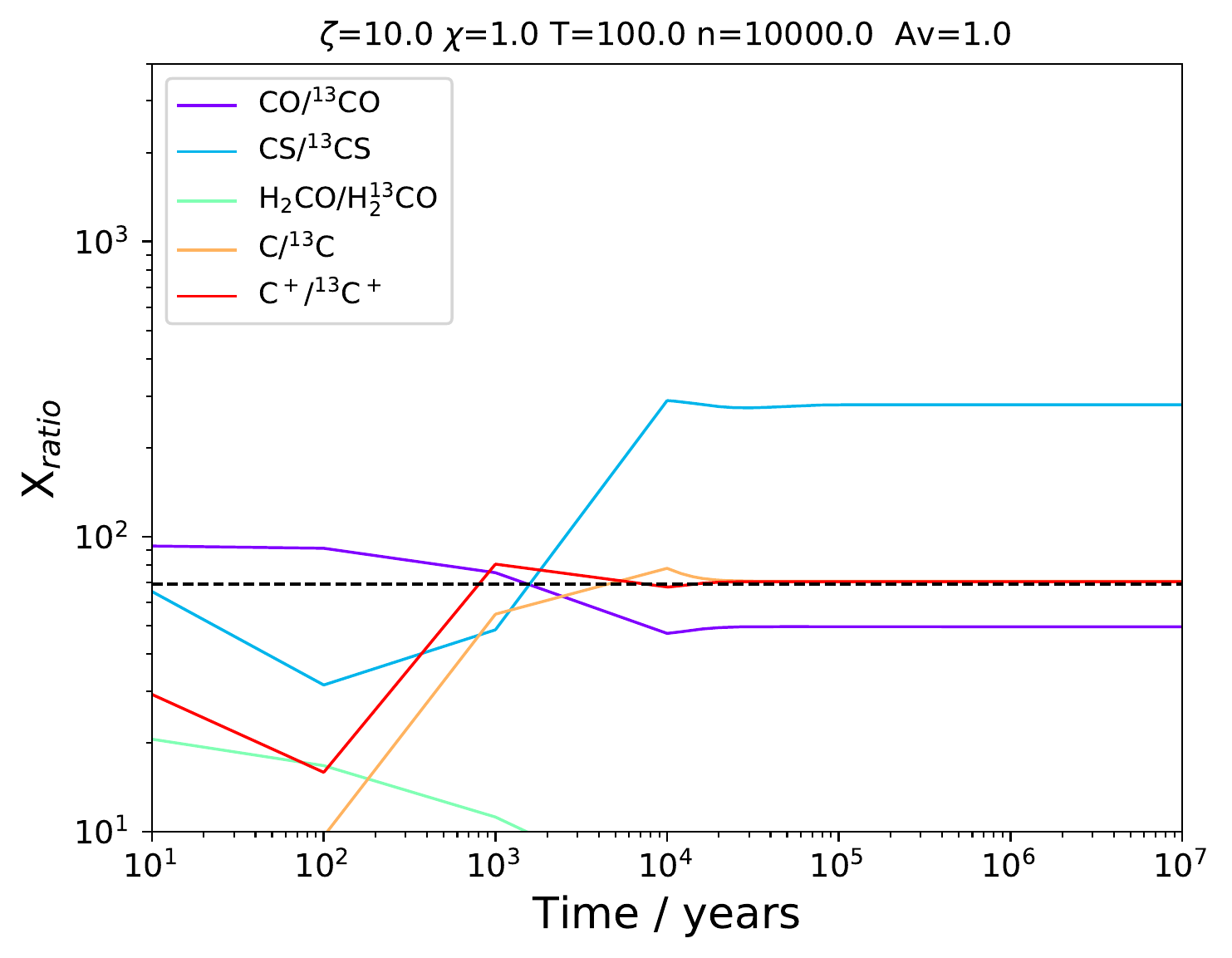}
    \caption{Examples of physical conditions under which  \Cratio in CO is lowest  (Top plot) or  higher (Bottom plot) than 68 (and for early times only) .}
    \label{fig:CO}
\end{figure} 

\item  For species other than CO, the carbon isotopic fractionation  varies with time and among molecules for most models, implying sensitivity to the several physical parameters we varied.  This extends the results by Roueff et al. (2015) where they also found, for a typical cold dense core, that the $^{12}$C/$^{13}$C isotopic ratios are very dependent on the evolution time. This indeed seems to be true even at higher temperatures when gas-grain interactions are no longer important. The isotopic ratio that seems to vary most across the parameters and with time is the one for the CS molecule. 

\item  Under standard galactic  dense clouds conditions, the  \Cratio ratio for CS, HCN and CN increases with time at very low extinctions (e.g. edge of the cloud), while it decreases with time deep into the clouds. The former behaviour is a consequence of photodissociation of the $^{13}$C molecular isotopologue into $^{13}$C and $^{13}$C$^+$ (note that the \Cratio ratio for C and C$^+$ does not vary because $^{13}$C and $^{13}$C$^+$ quickly react with CO and other species and 'return' to their main isotopologue form), while the latter is due to neutral-neutral reactions such as e.g. $^{13}$C + SO and $^{13}$C + CN. The behaviour seems to be reversed for HCO$^+$ and H$_2$CO (see Fig~\ref{fig:gal}). This result may be important depending on how much gas in a galaxy is in PDRs (low extinction) or e.g. in star forming regions (dense gas).

\begin{figure}
    \centering
    \includegraphics[scale=0.50]{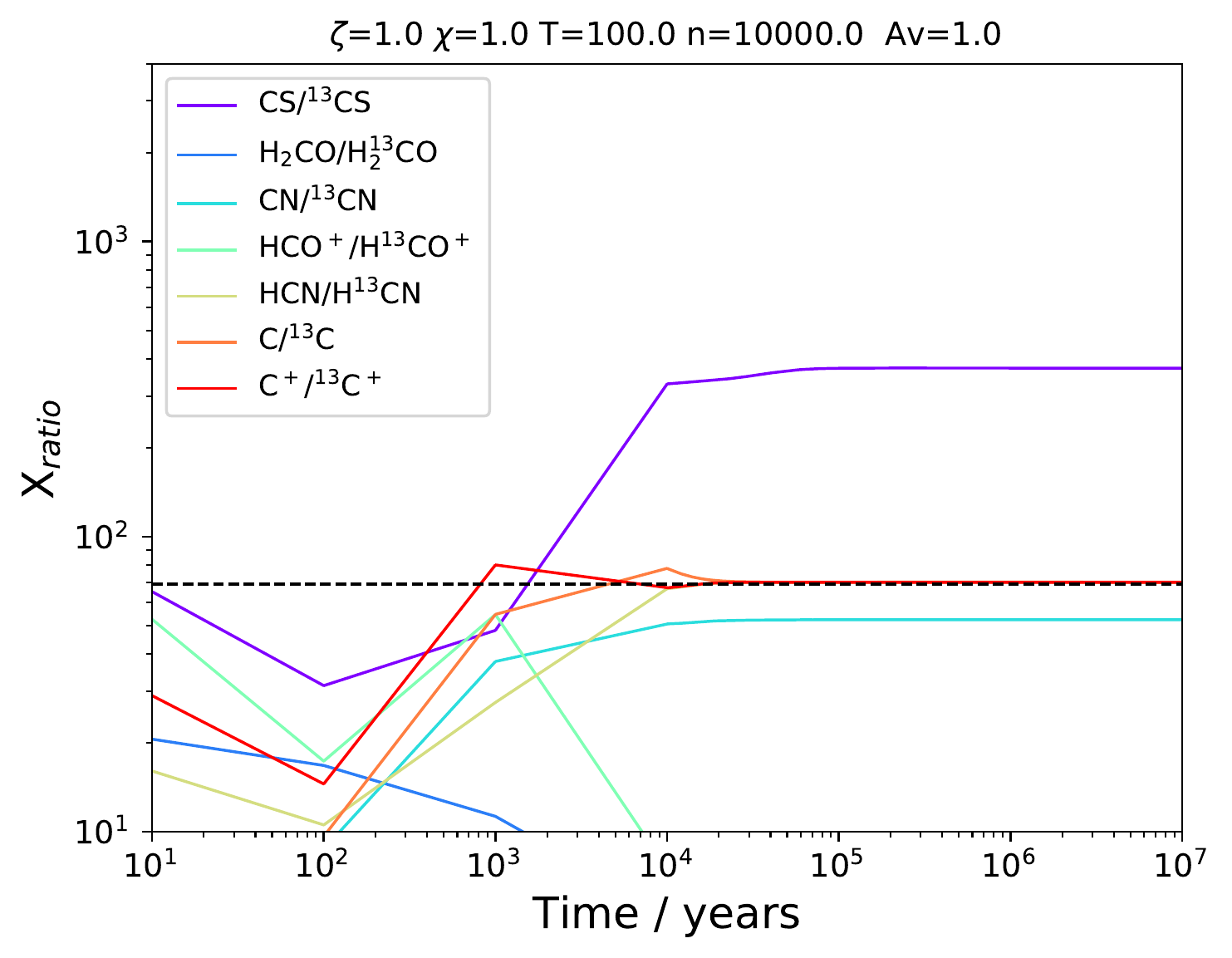}
    \includegraphics[scale=0.50]{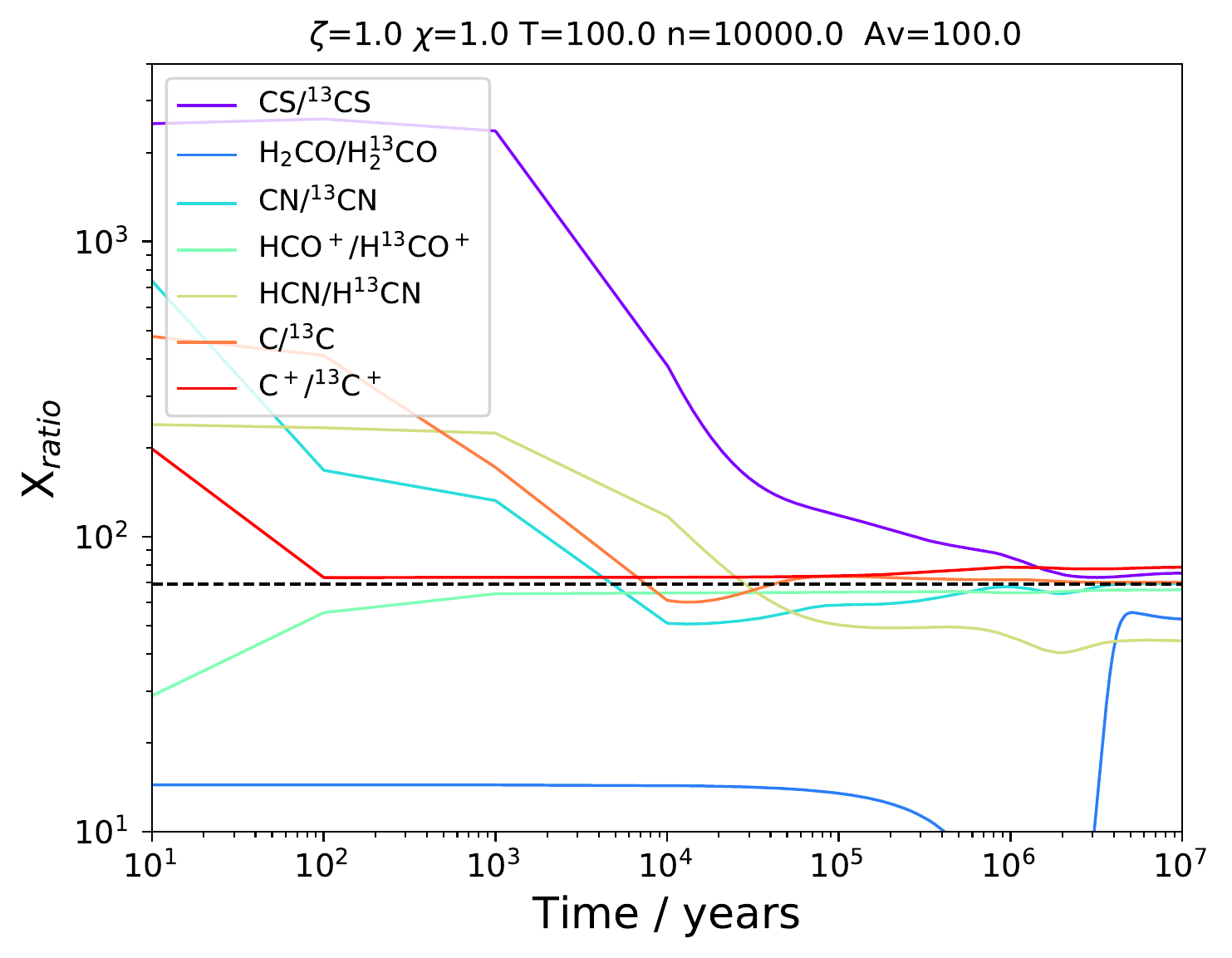}
    \caption{\Cratio for selected carbon-bearing species under typical dense cloud conditions at low visual magnitudes (Top) and high visual magnitudes (Bottom)}
    \label{fig:gal}
\end{figure} 

\item The effects of increasing the cosmic ray ionization rate varies across the physical parameters and are not linear with increasing $\zeta$. 
At low visual extinctions, there is an increase in the H$_2^{12}$CO/H$_2^{13}$CO, $^{12}$CS/$^{13}$CS, and H$^{12}$CN/H$^{13}$CN ratio especially at early times and most pronounced for CS. This is mainly due to photodissociation, either via direct photons or by secondary photons. At higher visual extinctions, and early times, an increase in $\zeta$ leads to a decrease and an increase in the \Cratio for HCO$^+$ and CS respectively. At high densities steady state is reached very early (in $<$ than 100 years) and, at low A$_V$, the only effect of increasing $\zeta$ seems to be a decrease in the \Cratio ratio for H$_2$CO, while at high A$_V$, the \Cratio ration for HCN slightly increases while it heavily decreases for HCO$^+$ at very high cosmic ray ionization rates. We show an example of the effect of increasing comic ray ionization rates in Figure~\ref{fig:highzeta} which can also be compared with Figure~\ref{fig:gal}.

\begin{figure}
    \centering
    \includegraphics[scale=0.50]{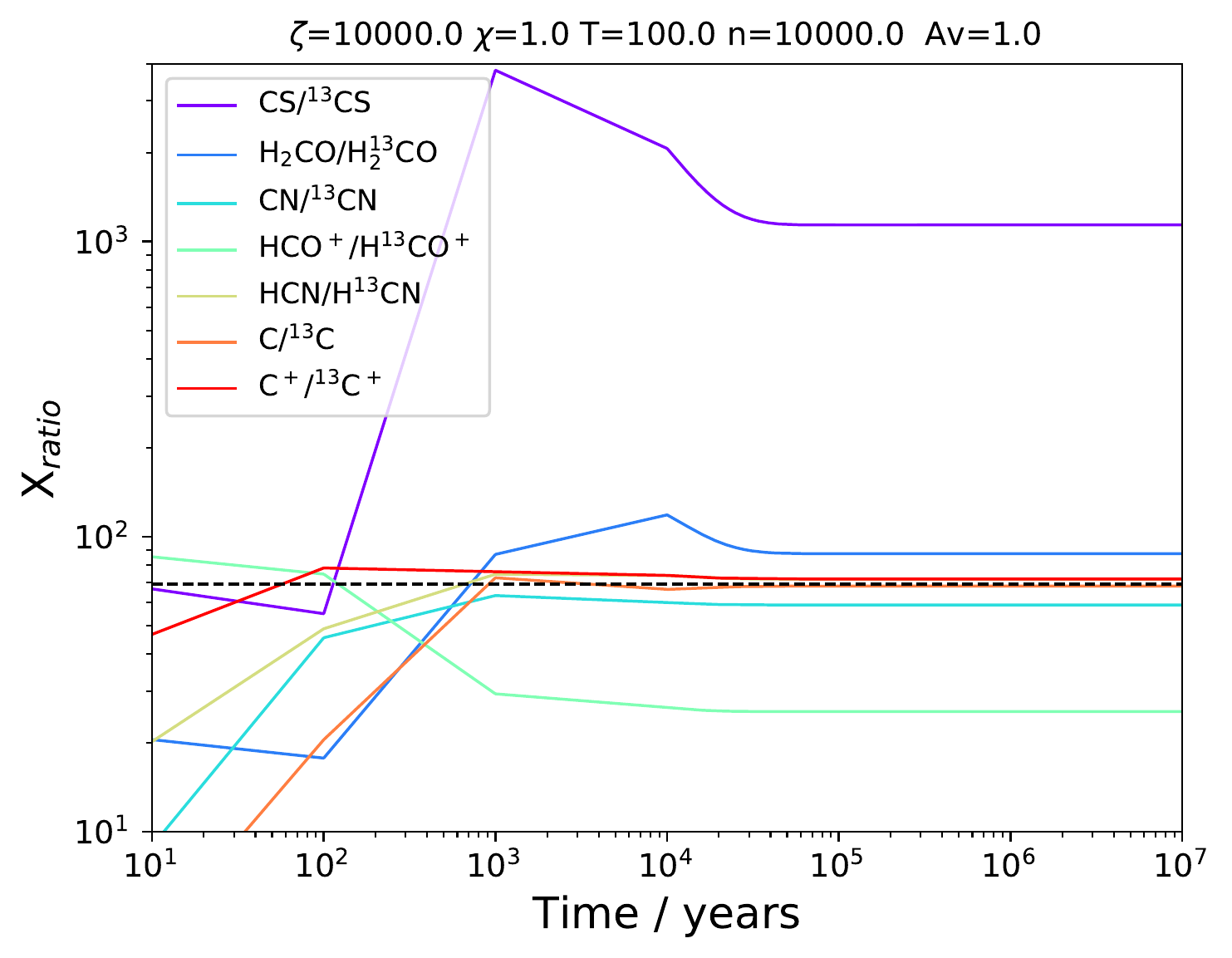}
    \includegraphics[scale=0.50]{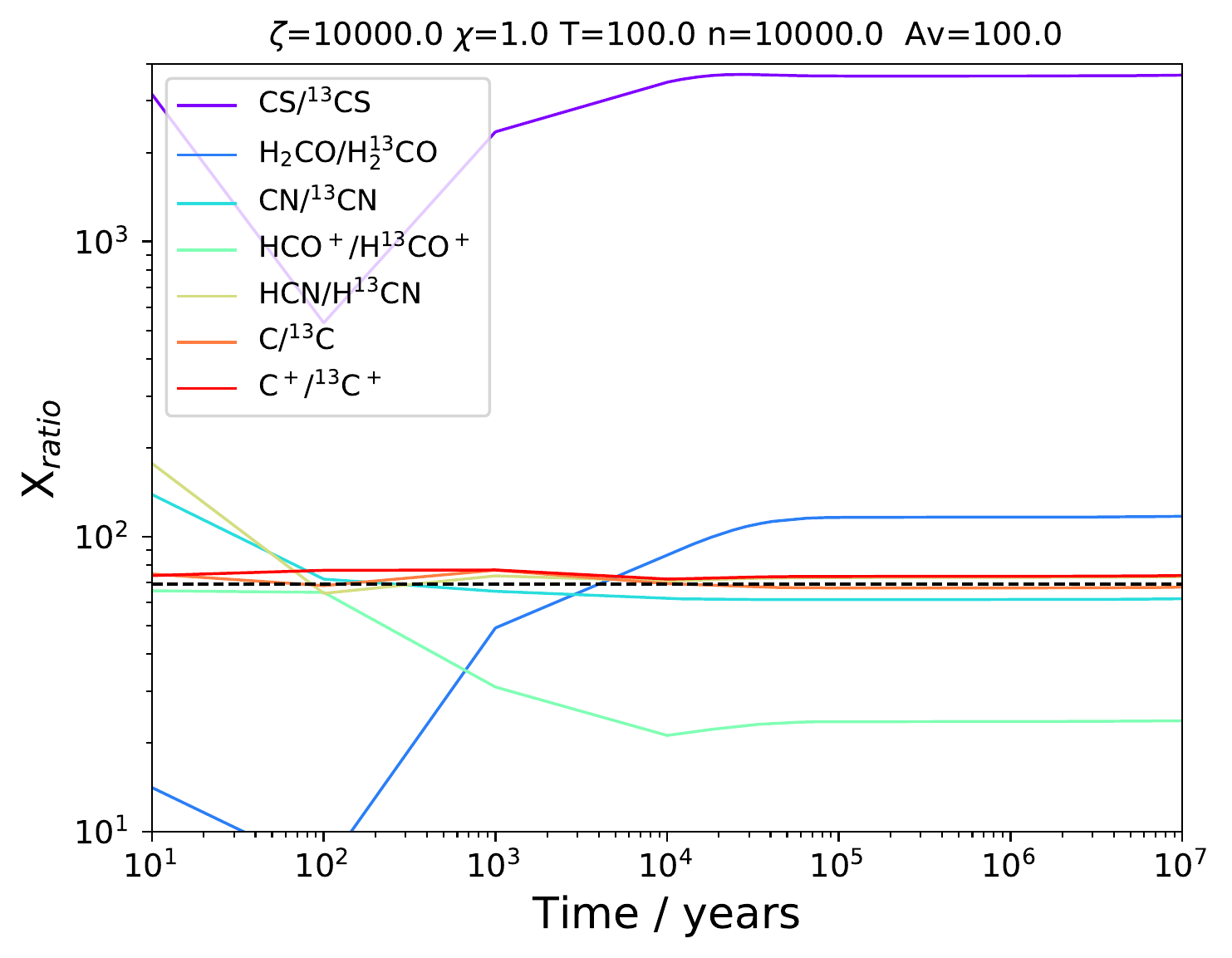}
    \includegraphics[scale=0.50]{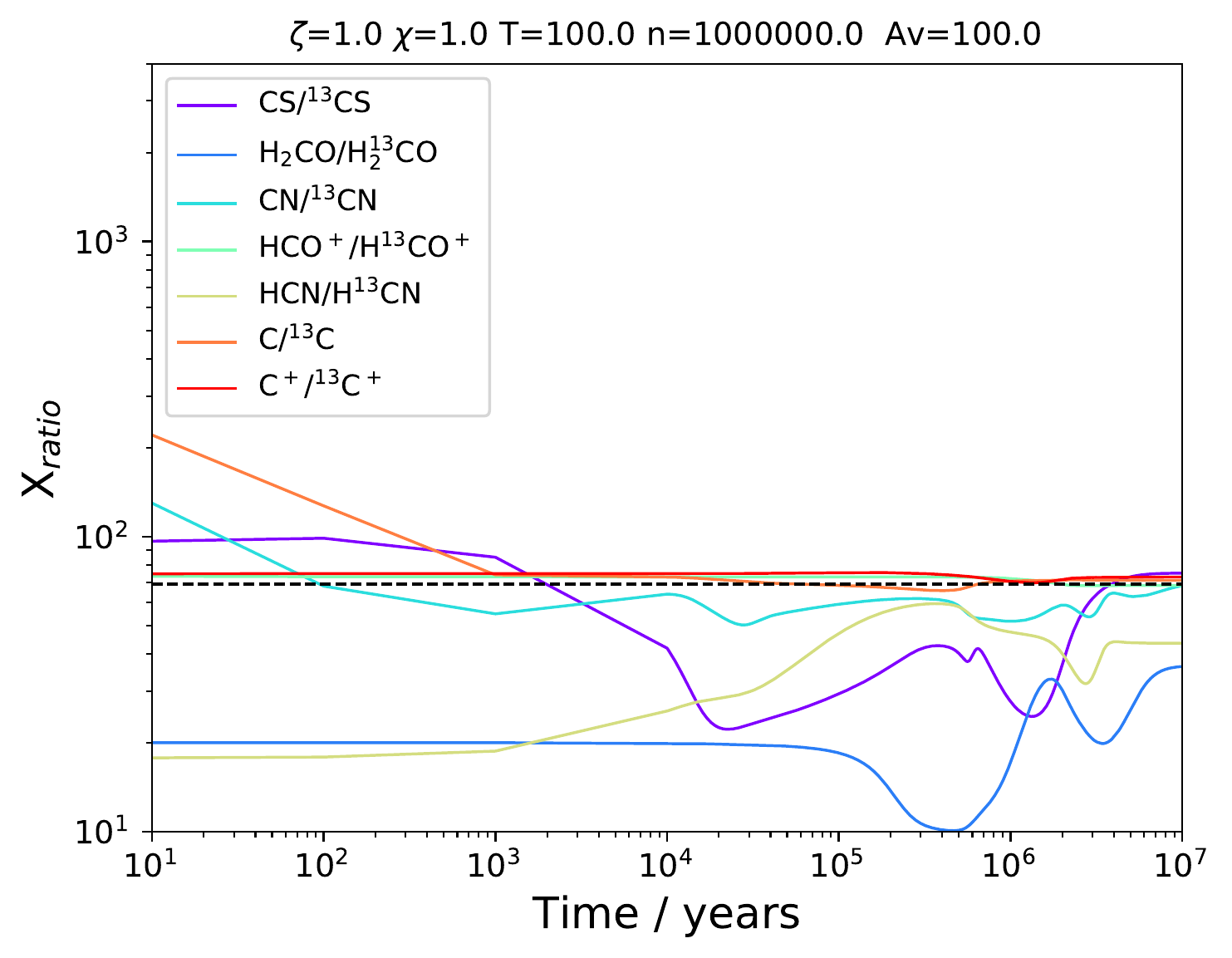}
    \includegraphics[scale=0.50]{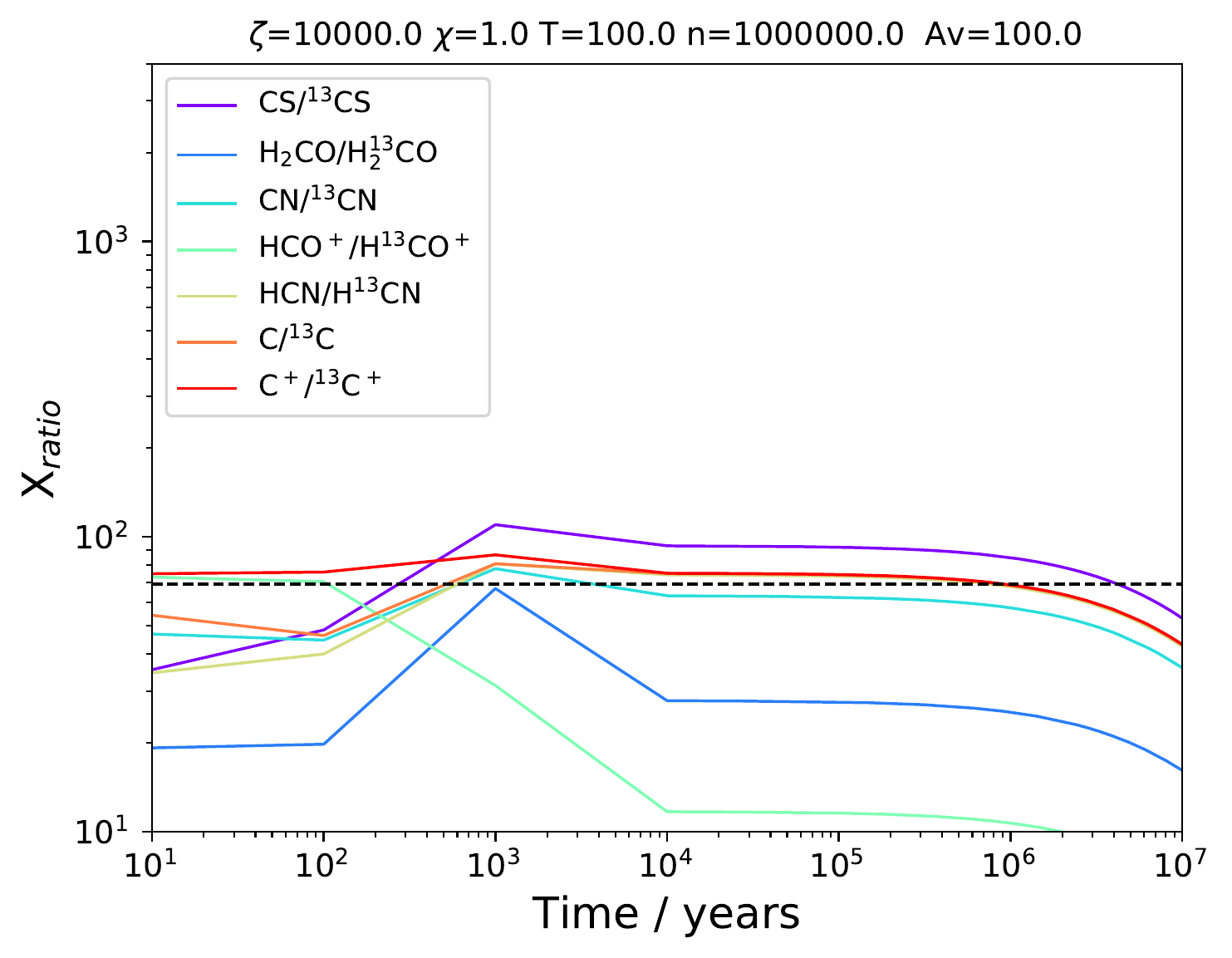}
    \caption{\Cratio for selected species and models showing the effects of varying the cosmic ray ionization rate. }
    \label{fig:highzeta}
\end{figure} 
\item  An increase in  radiation field strength only has effects at low magnitudes. At low densities it leads to a larger variation in time for CS and an increase in \Cratio ratio for both H$_2$CO and HCO$^+$. At high densities, we see a decrease followed by an increase in \Cratio ratio for HCO$+$, as well as a large increase for the \Cratio ratio of CS.
In general a higher radiation field leads to an earlier steady state value for the \Cratio ratio. An example can be seen in Figure~\ref{fig:highrad}.
\begin{figure}
    \centering
    \includegraphics[scale=0.50]{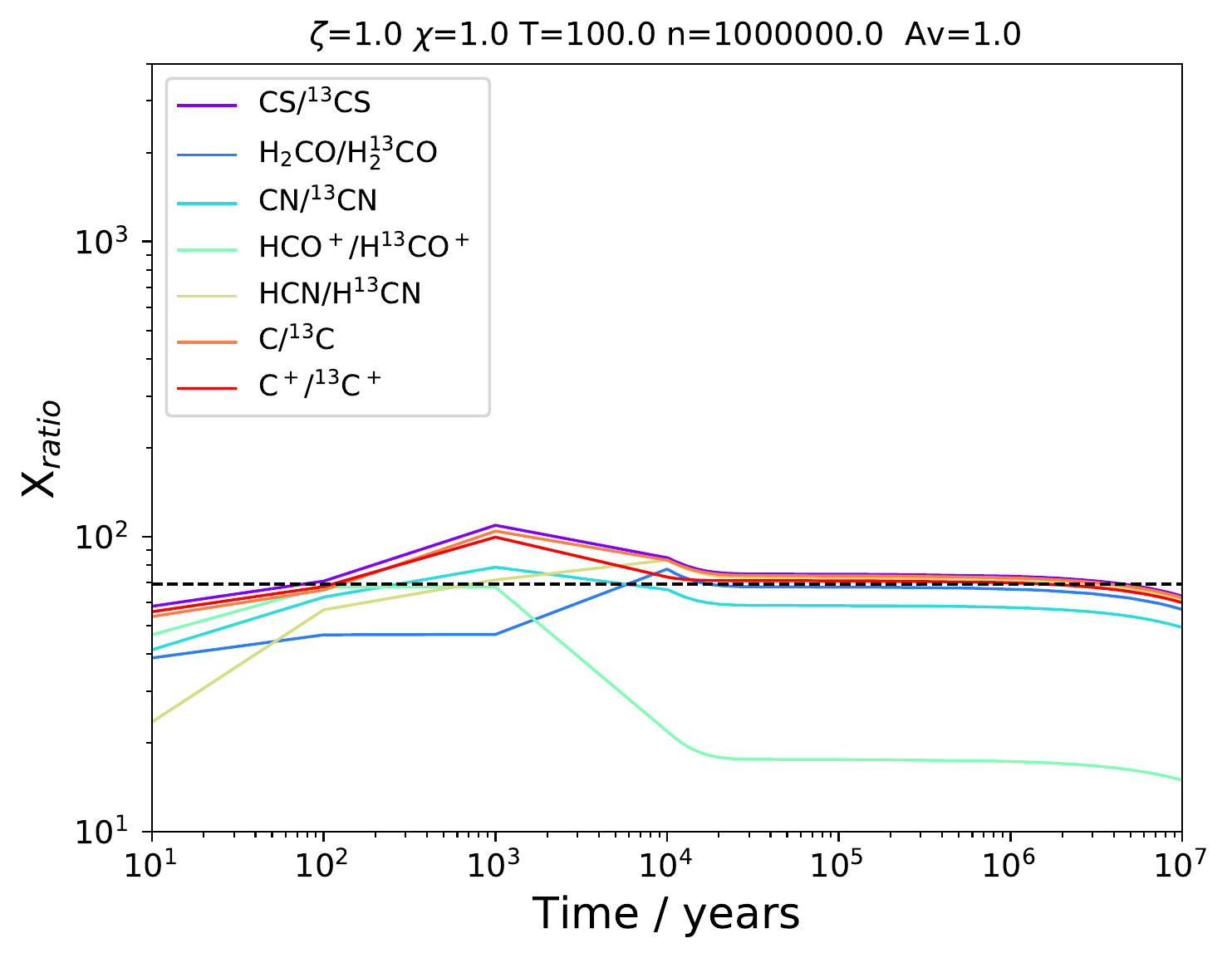}
    \includegraphics[scale=0.50]{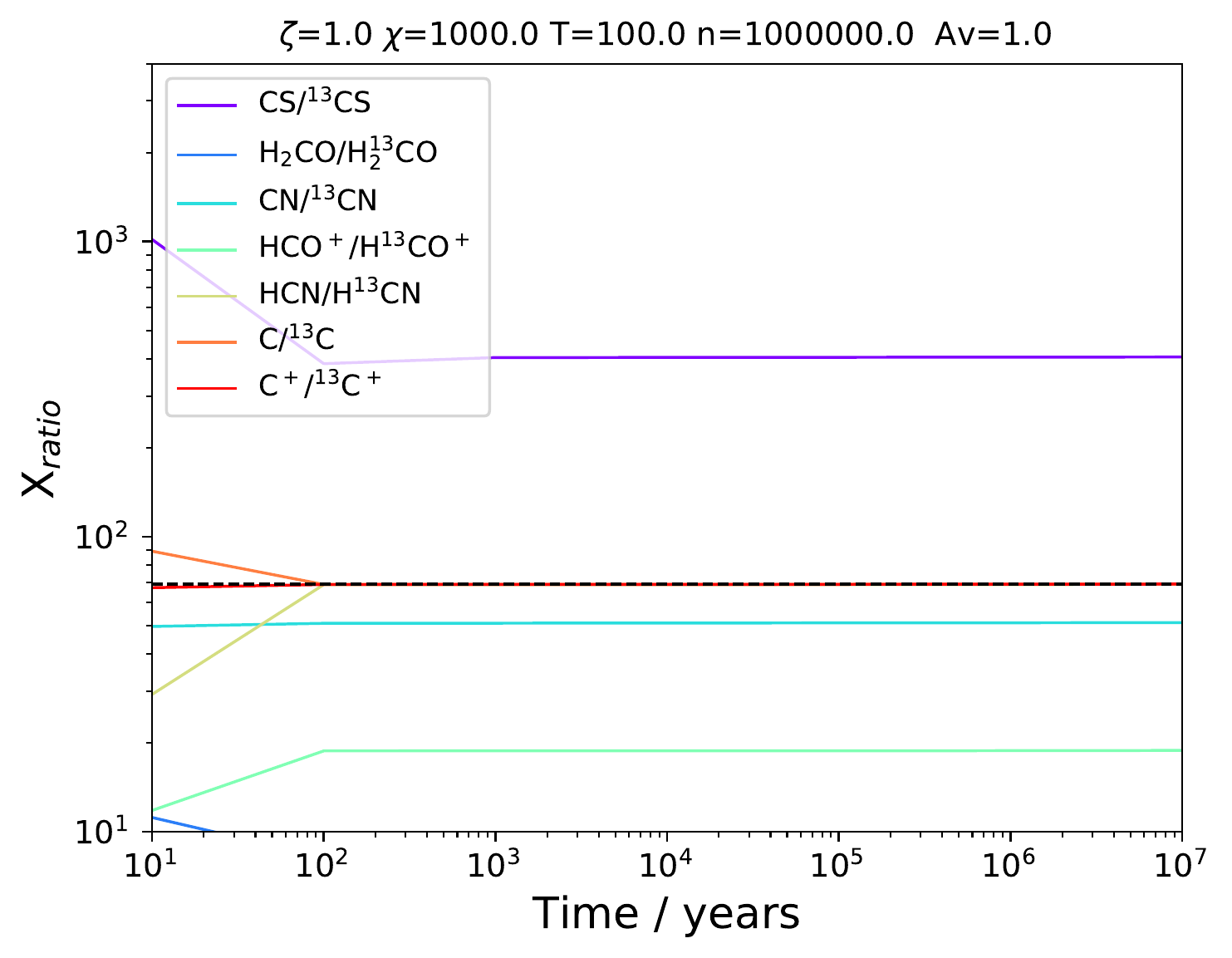}
    
    \caption{\Cratio for selected species and models showing the effects of varying the radiation field strength. }
    \label{fig:highrad}
\end{figure} 
\item In environments where both radiation field and cosmic ray ionization rates are enhanced, the $^{12}$C/$^{13}$C ratio,  as a function of time and density, differs across molecules  very non linearly.   As we shall see in Section 3.3, this may be able to explain the large spread found across observations.
\begin{figure*}
    \centering
    \includegraphics[scale=0.42]{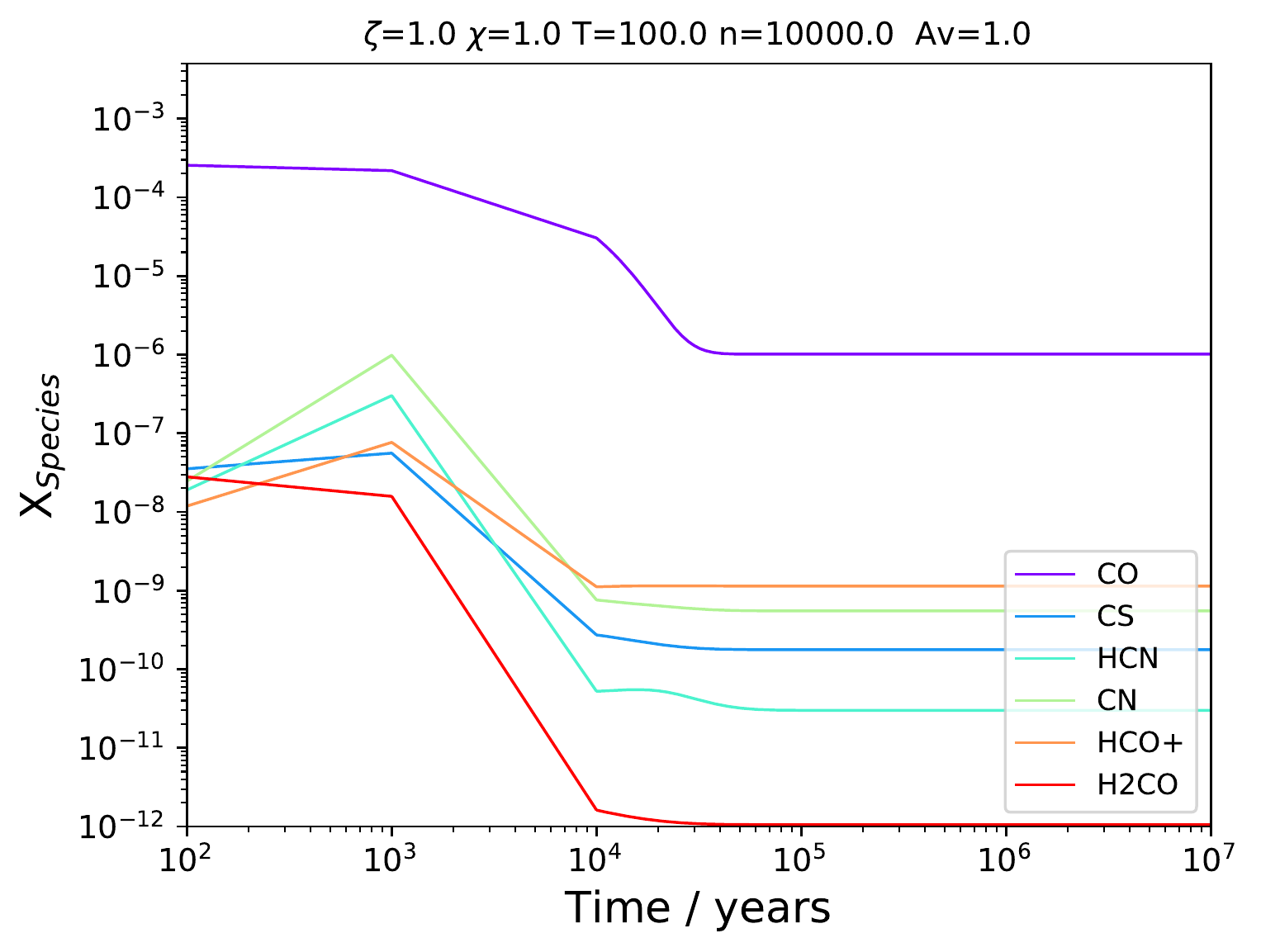}
    \includegraphics[scale=0.42]{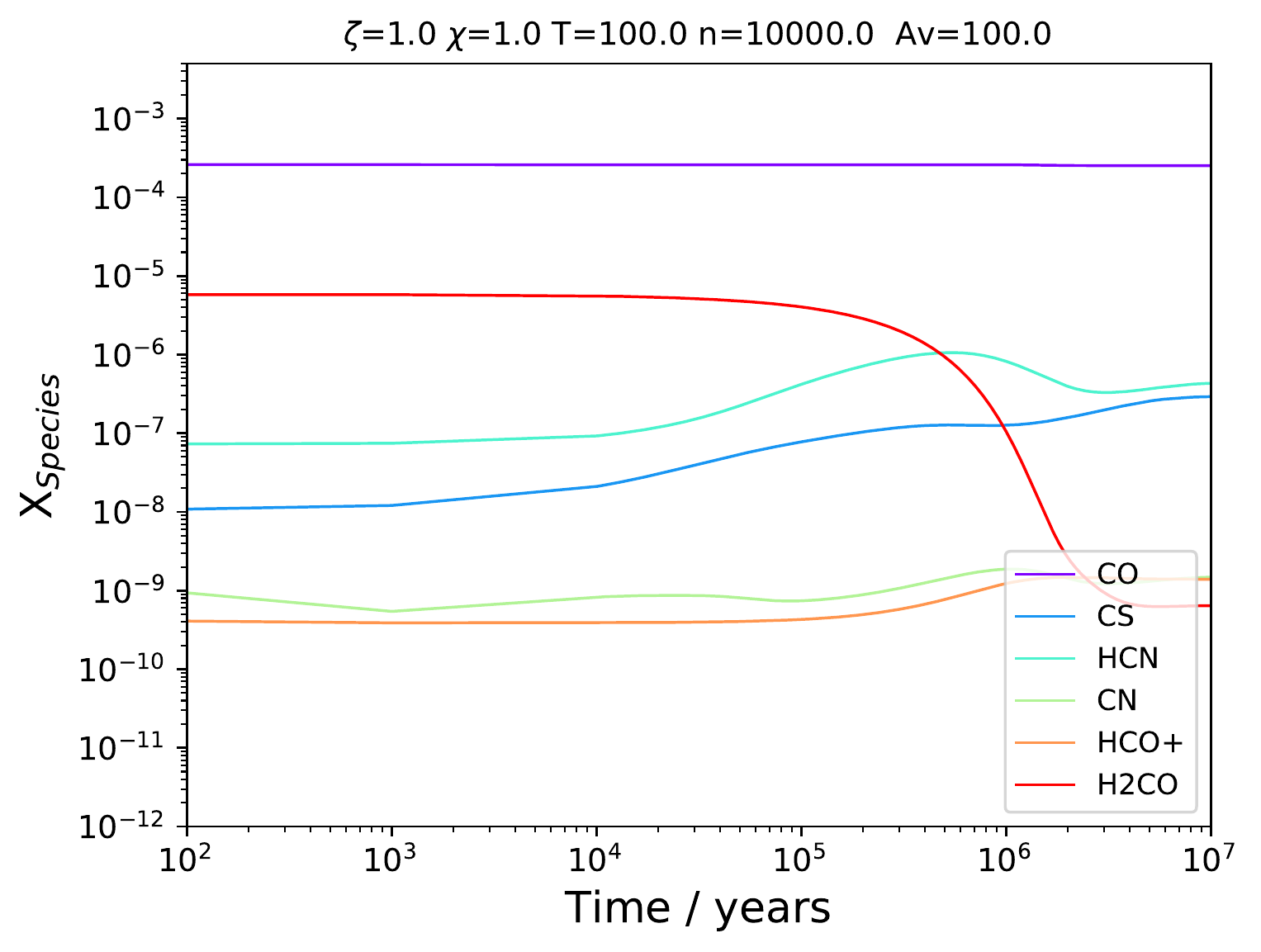}
    \includegraphics[scale=0.42]{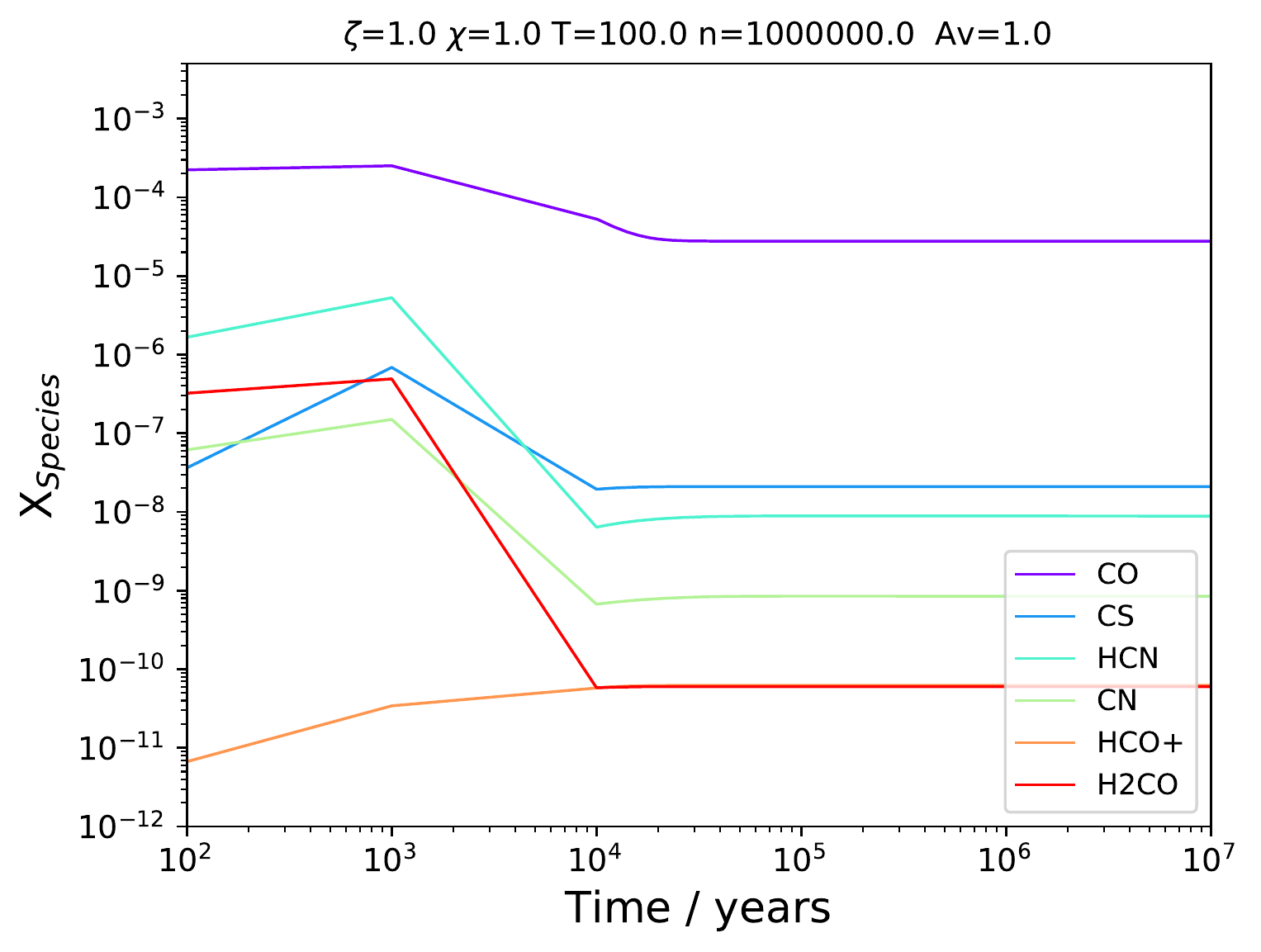}
    \includegraphics[scale=0.42]{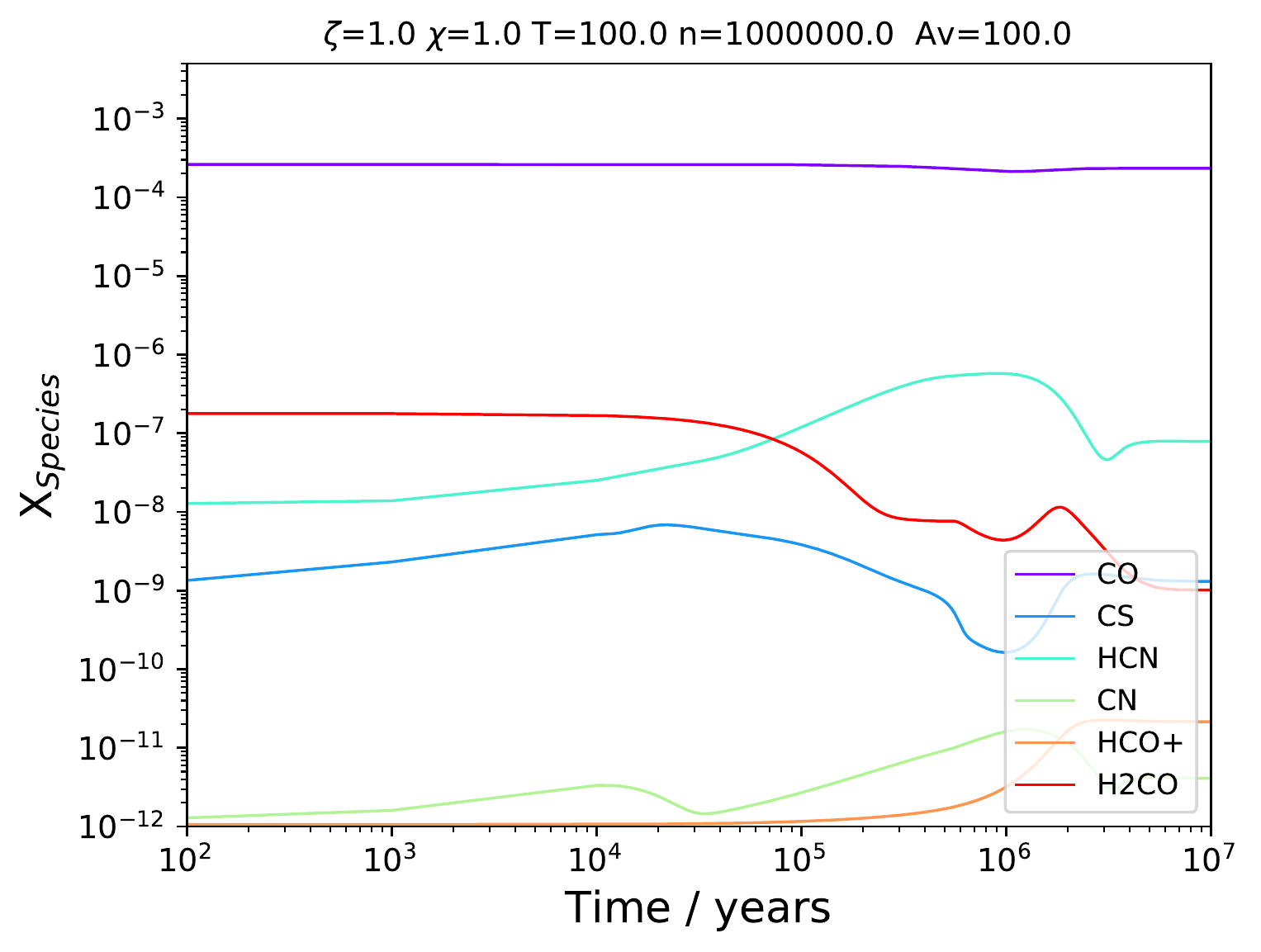}
    \includegraphics[scale=0.42]{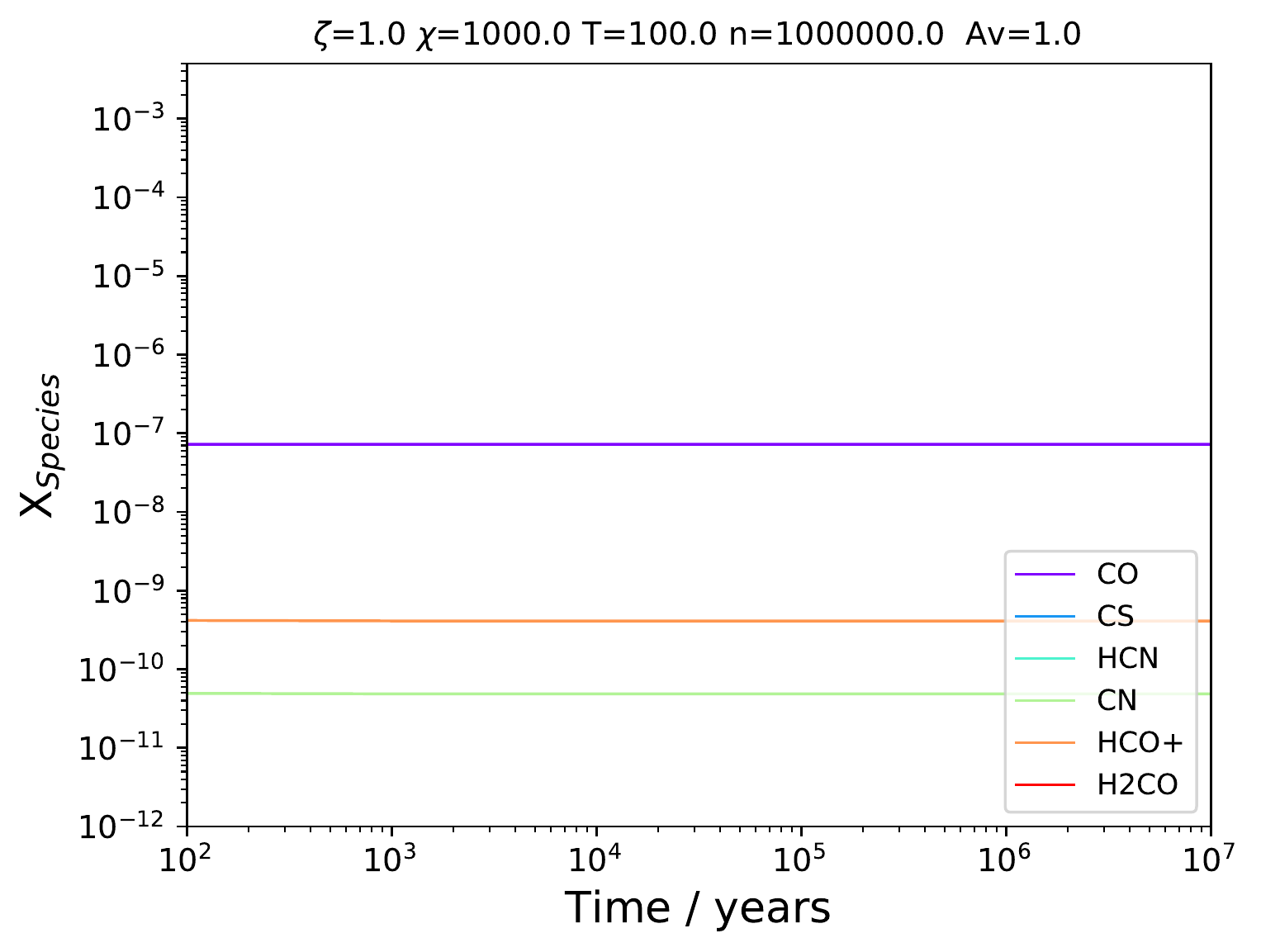}
    \includegraphics[scale=0.42]{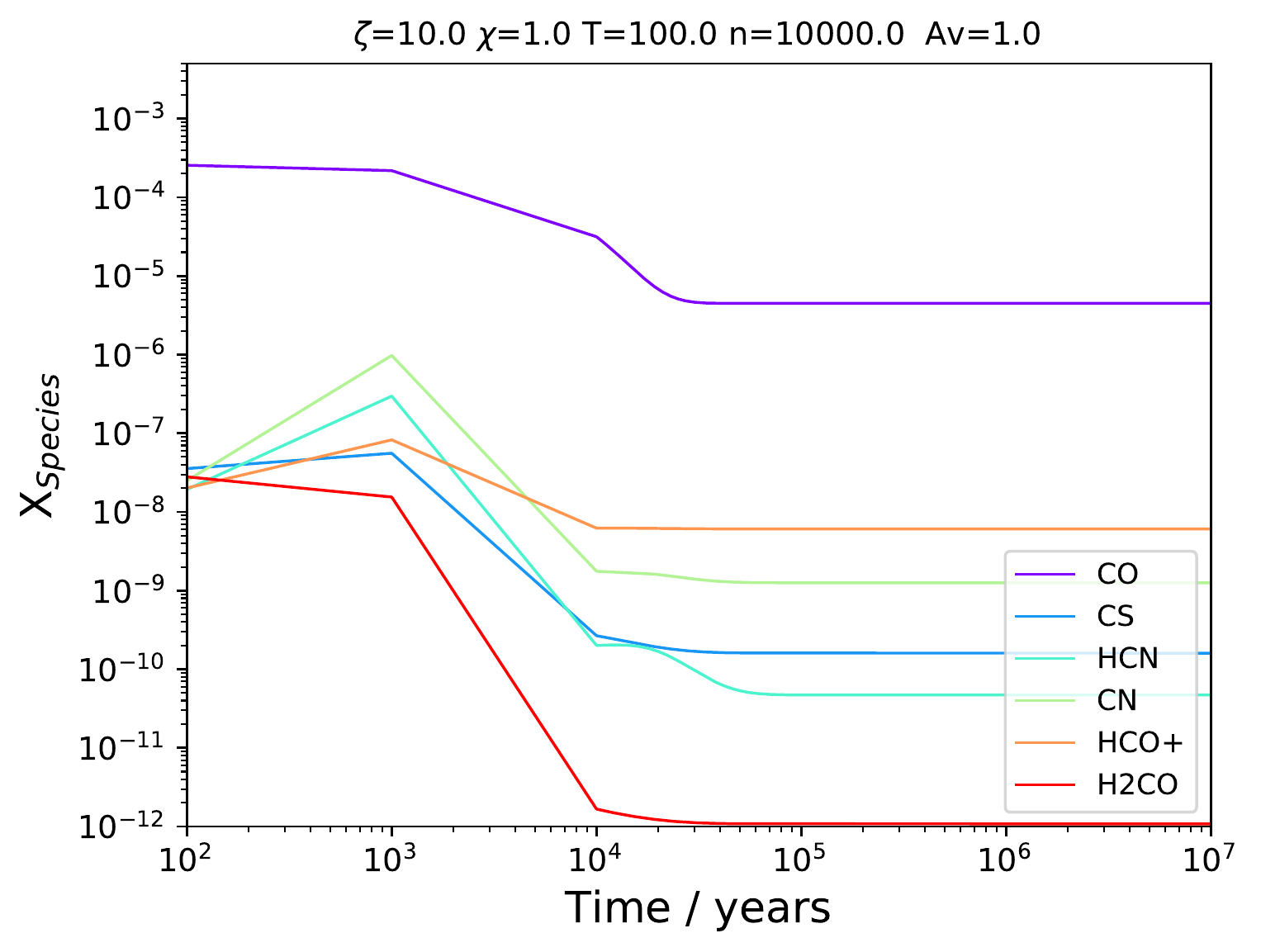}
    \includegraphics[scale=0.42]{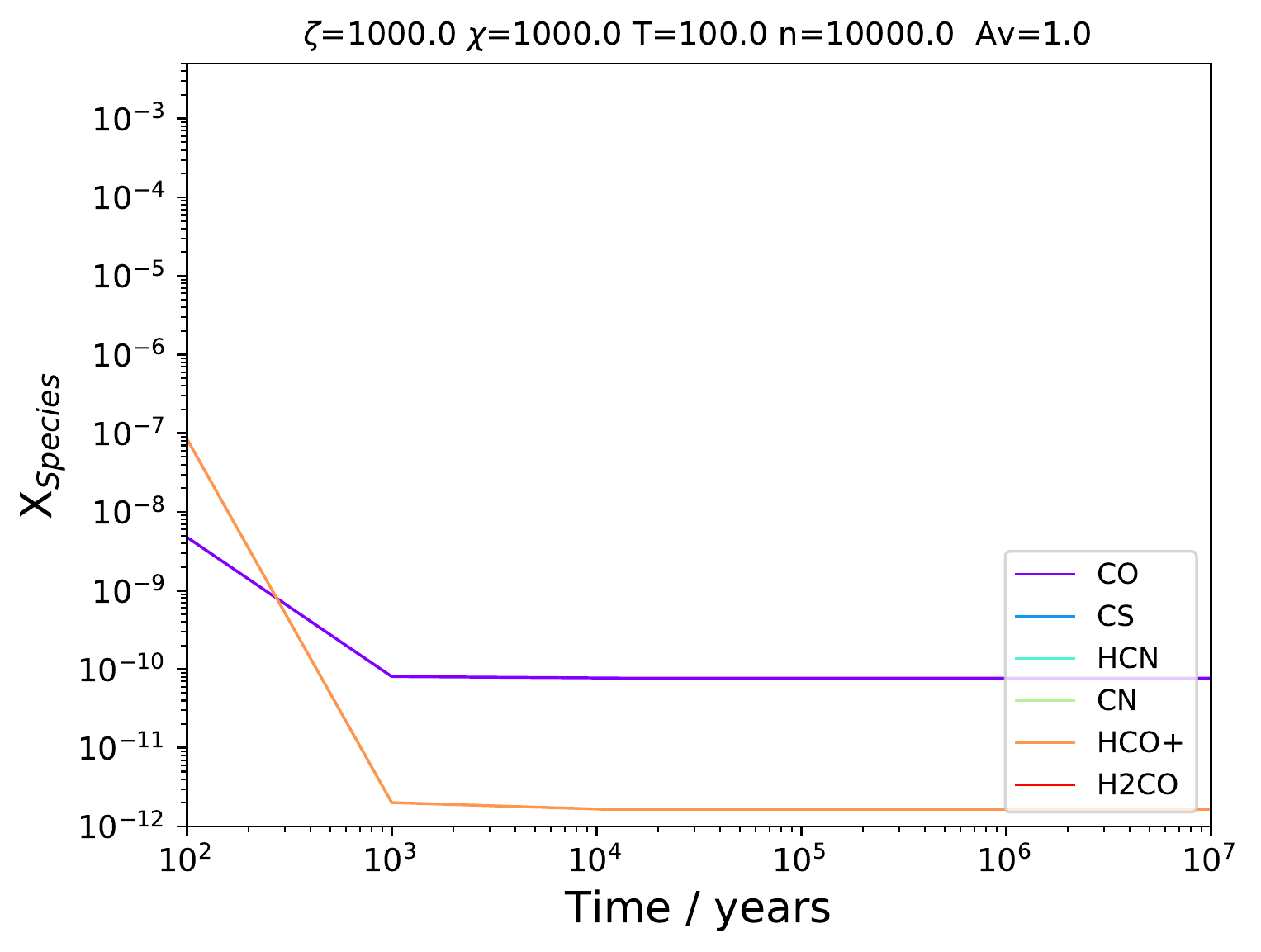}
    \includegraphics[scale=0.42]{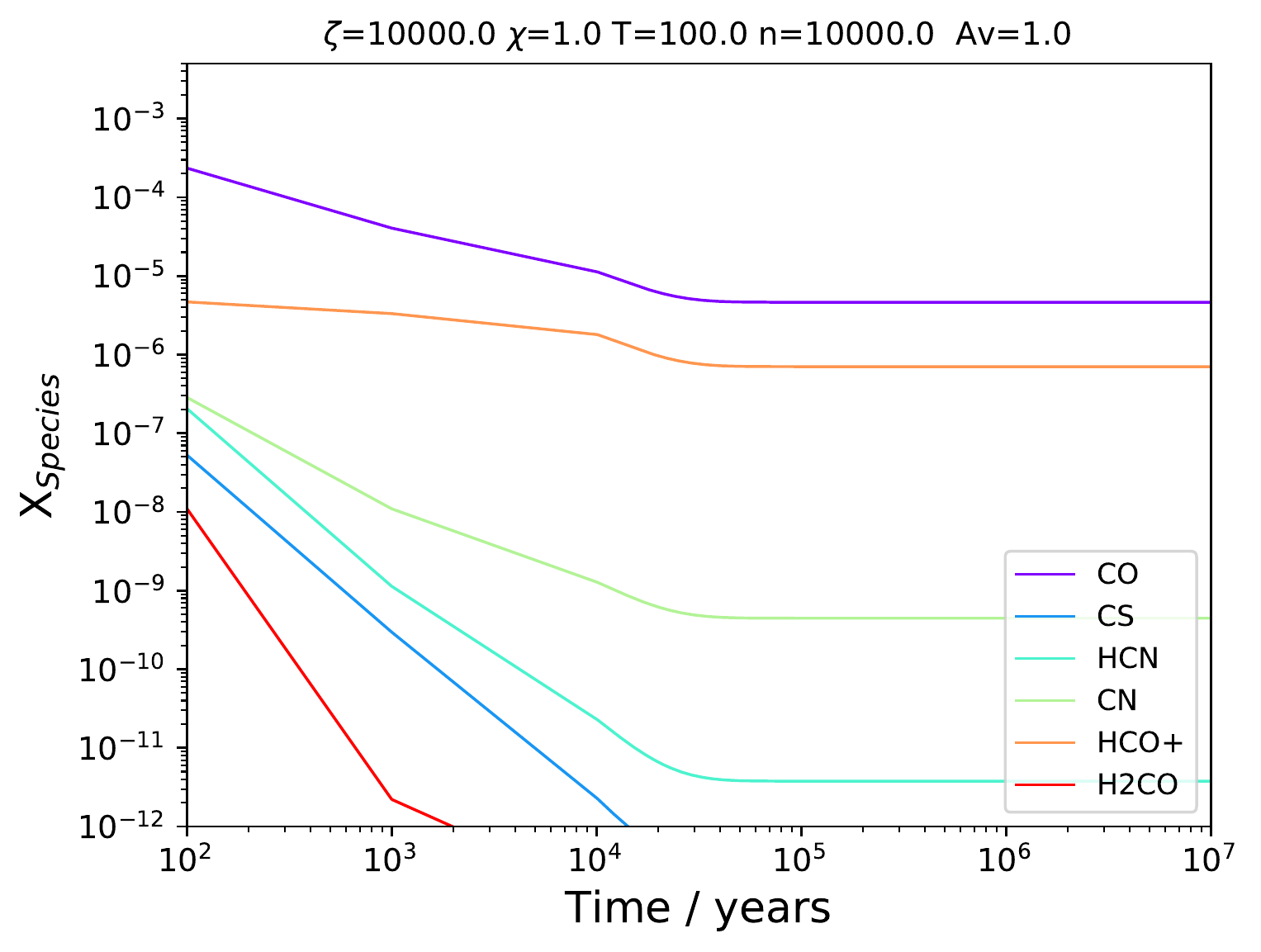}
    \includegraphics[scale=0.42]{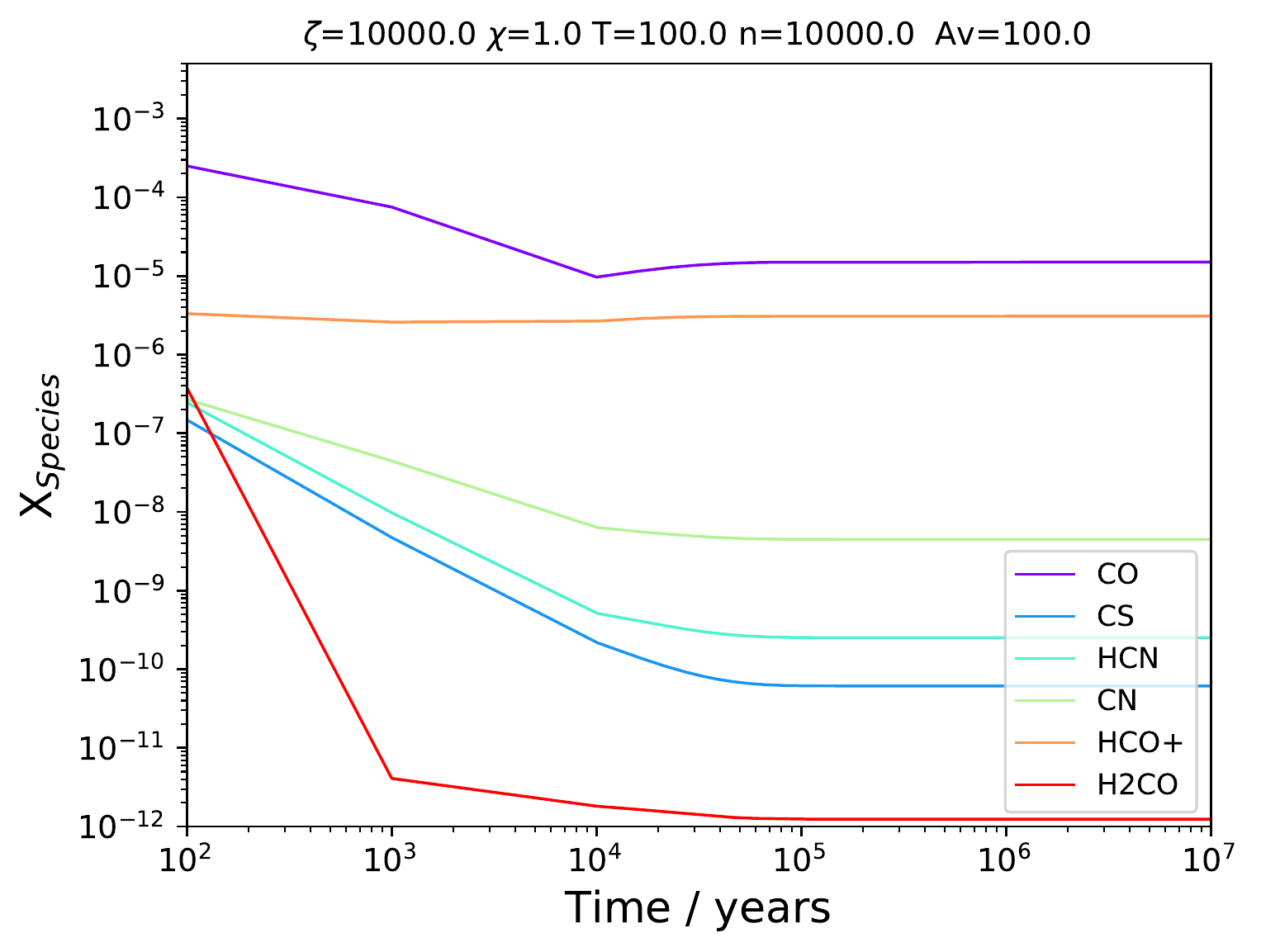}
    \includegraphics[scale=0.42]{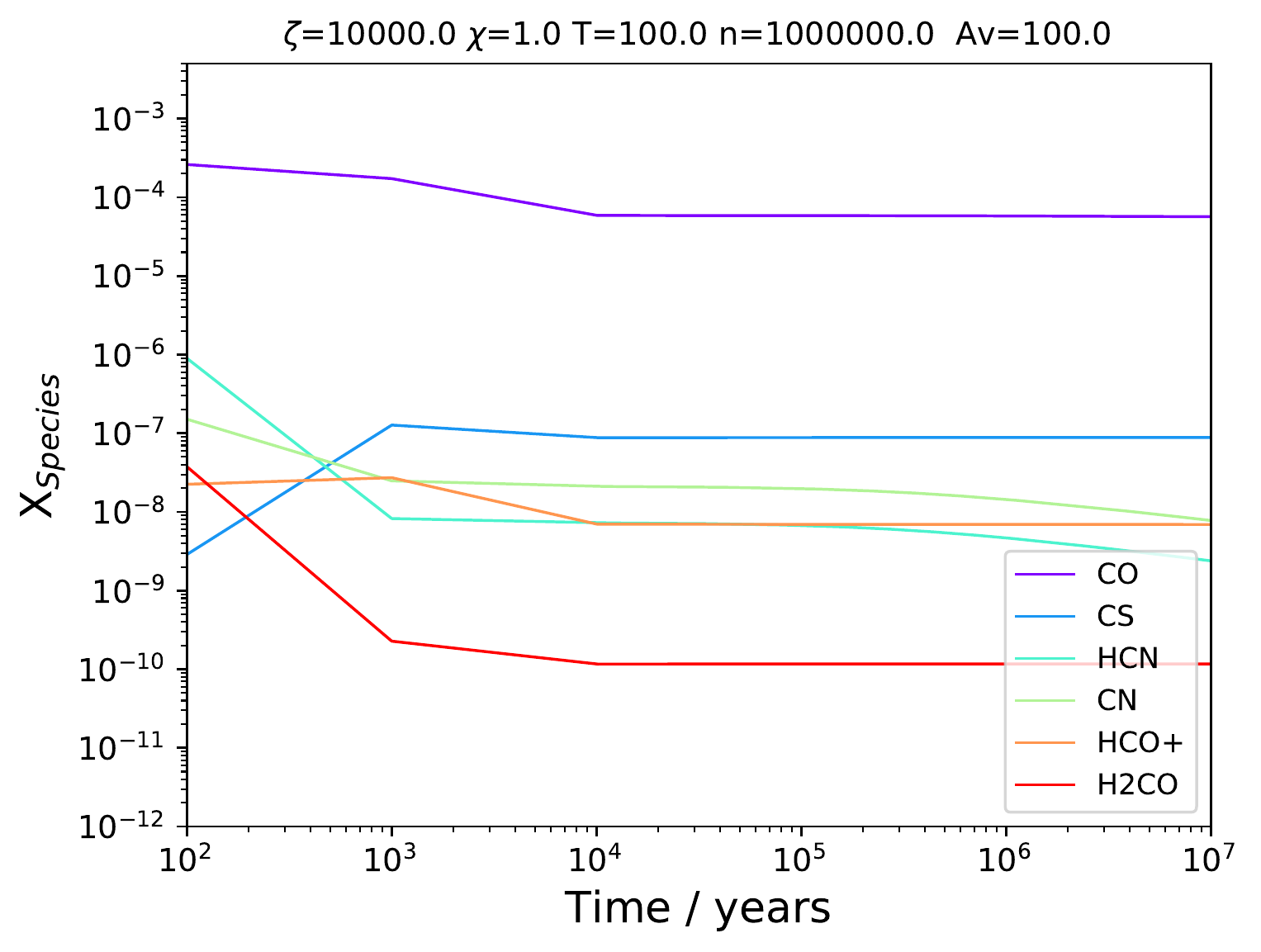}
   \caption{Fractional abundances of the species for which their isotopic ratios are discussed in Section 3.1.}
   \label{fig:abundances}
\end{figure*}

A summary of the qualitative trends of the $^{12}$C/$^{13}$C ratio with time, as a function of the combination of the physical parameters we varied in our models, is given in Table~\ref{tb:variations}. 
\end{itemize}
\begin{table*}
        \centering
        \caption{Qualitative trends of the $^{12}$C/$^{13}$C ratio as a function of different parameters for species other than CO.}
        \label{tb:variations}
        \begin{tabular}{lcccc}
                \hline
                Model & \multicolumn{2}{c}{$A_{\rm V}=1$ mag} &  \multicolumn{2}{c}{$A_{\rm V}\geq 10$ mag} \\
                \hline
                standard & \multicolumn{2}{c}{Increases  with time at low density only}  &  \multicolumn{2}{c}{Decreases  with time  at low density only } \\
        high $\zeta$ & \multicolumn{2}{c}{Increase with time at low $n_H$} & \multicolumn{2}{c}{Decrease/Increase at early times } \\
         & \multicolumn{2}{c}{At high $n_H$, very early steady state} & \multicolumn{2}{c}{HCO$^+$/CS (low $n_H$ only)} \\
                high $\chi$  &  \multicolumn{2}{c}{Increase for H$_2$CO and HCO$^+$ at low $n_H$}  & \multicolumn{2}{c}{No sensitivity to $\chi$} \\
              & \multicolumn{2}{c}{Increase for CS at high $n_H$.}  \\
        high $\zeta + \chi$ & \multicolumn{2}{c}{Highly variable, especially for CS} & \multicolumn{2}{c}{Highly variable, especially for CS} \\
                \hline
        \end{tabular}
    \label{variations}
\end{table*}

 The complex, highly time and parameter dependent, behaviour of the \Cratio ratio is a consequence of the many different reaction channels involved in the $^{13}$C chemistry (although we again note that our network only includes singly fractionated molecules). 

\subsection{Differences in fractionation among C-bearing molecules}
\label{diff_molecules}

Differentiation in fractionation of carbon has been investigated by other authors before, but usually at temperatures of $\leq$ 50 K  (Roueff et al. (2015); Colzi et al. (2020)). For example, Roueff et al. (2015) find that, in a model ran at both 2$\times$10$^4$ (a) and 2$\times$10$^5$ cm$^{-3}$ (b) at 10 K, the $^{12}$C/$^{13}$C is much higher in HCN than in CN at most times during the chemical evolution of the gas.  As explained in Section 2, in order to test our network, we ran a model similar to model (b) presented in Roueff e al. (2015) and find indeed the same trend at 10~K. 

Within our grid of models, we find that there are always large variations among most of the carbon bearing species we looked at. In order to qualitatively summarise these differences in the context of the environments we modelled, we shall divide our discussion according to different combination of physical and energetic parameters qualitatively mimicking different types of galaxies. As a caveat to this approach, we of course note that assigning particular characteristics to a class of galaxies is by definition an oversimplification. For this Section, we shall only consider models up to densities of  10$^5$ cm$^{-3}$ as, observationally, in most cases the beam is unlikely to encompass {\it average} densities higher than that.

\begin{figure}
   \includegraphics[scale=0.55]{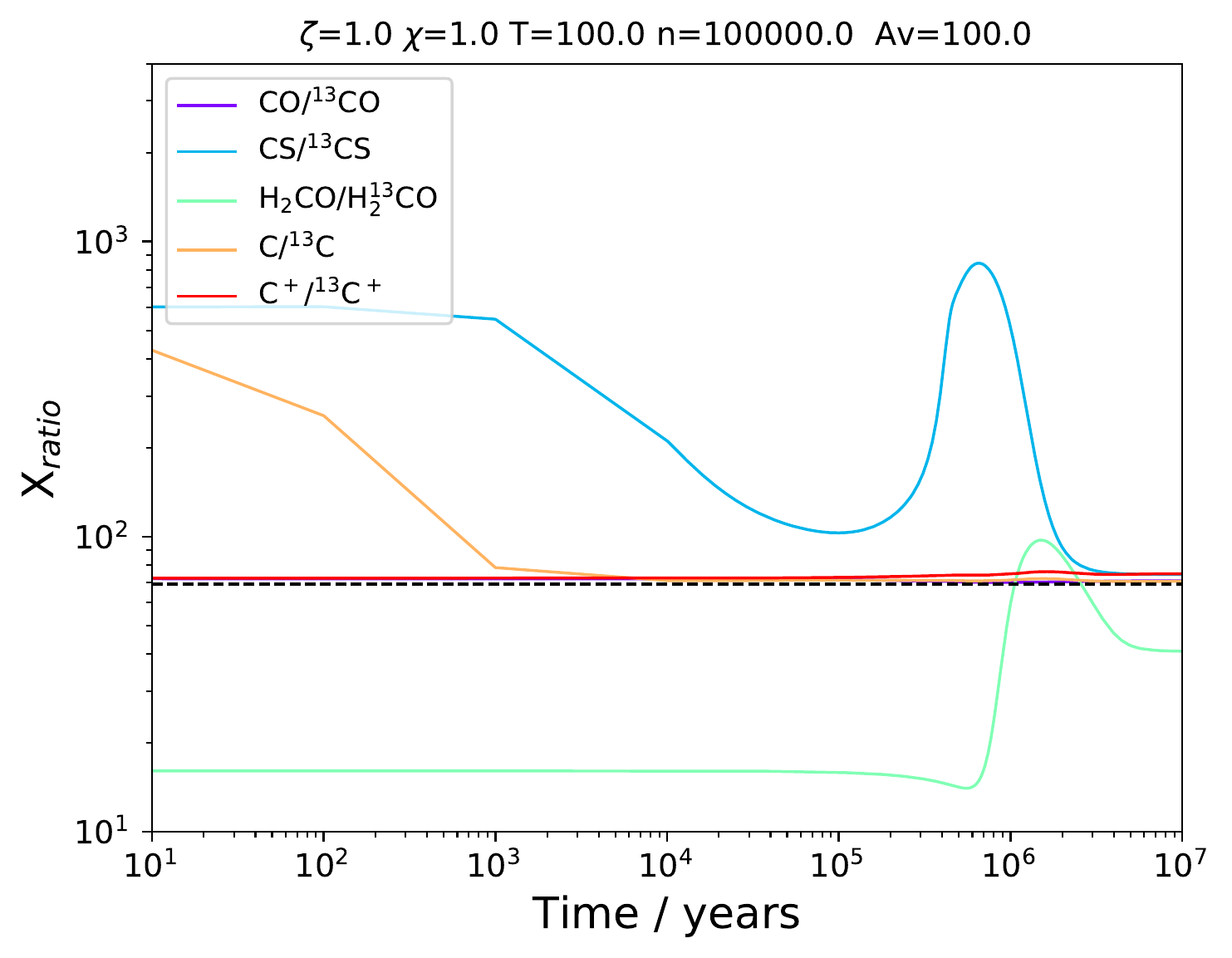}
   \includegraphics[scale=0.55]{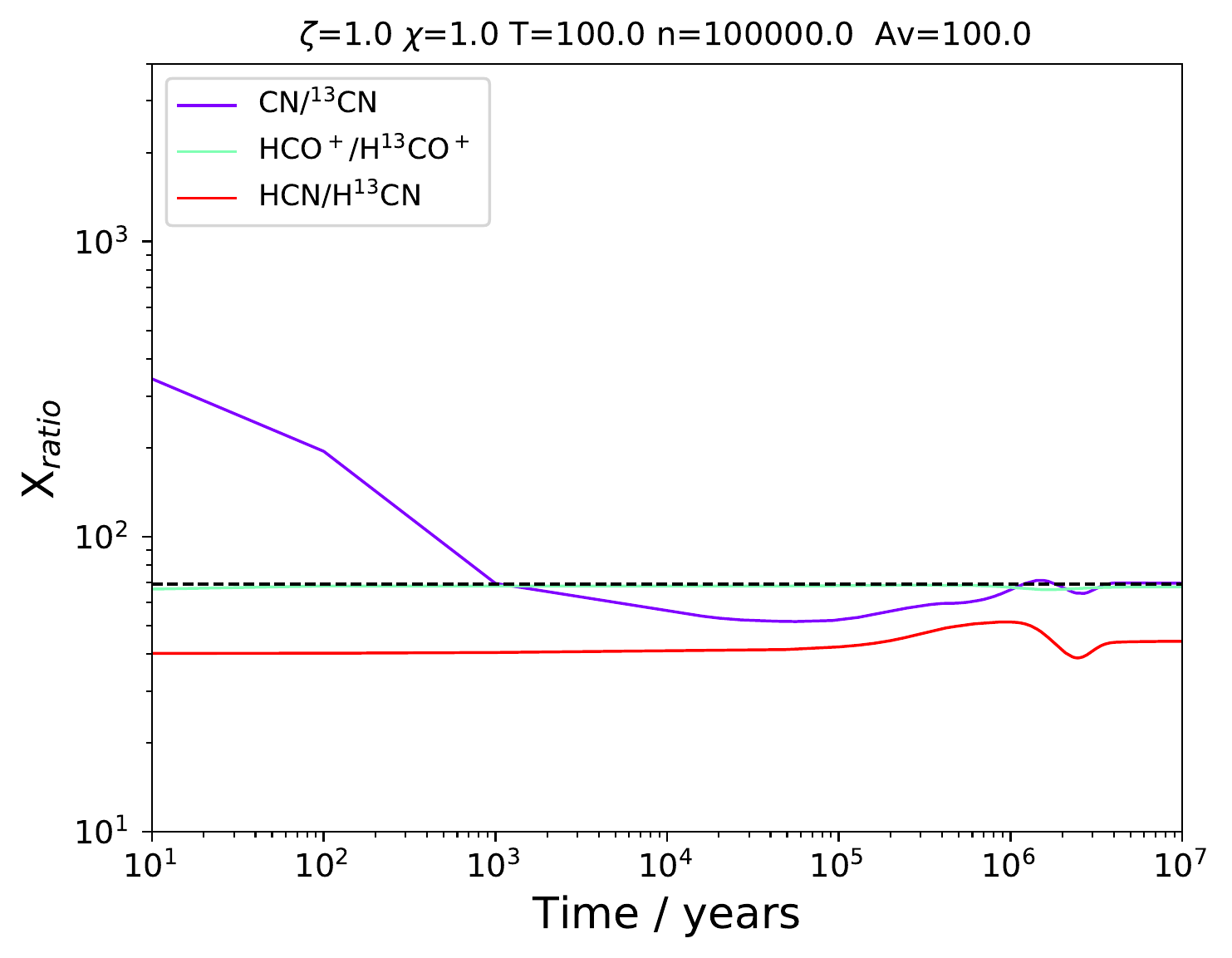}
\caption{\Cratio for selected species in spiral galaxies at high densities.}
\label{fig:denseGMC}
\end{figure}

{\it Spiral Galaxies}: assuming these galaxies are similar to the Milky Way from an energetic point of view (and hence we can model them with a standard cosmic ray ionization rate and radiation field strength), we find that  the large scale  molecular gas (with gas density of 10$^4$ cm$^{-3}$ and an A$_V$ = 1 mags), has a rather constant \Cratio for CO  of $\sim$ 69  but we see other species vary within a range going from $\leq$20 (for H$_2$CO and HCO$^+$) to 300 or so (for CS).   By  10,000 years, steady state is reached and the only species that have a \Cratio ratio close to the canonical one of $\sim$ 70 is HCN, with the CN \Cratio ratio slightly lower (60). (cf Fig~\ref{fig:gal} Top). 
For higher  visual extinction,  still at densities of 10$^4$ cm$^{-3}$, (mimicking possibly the gas in Giant Molecular Clouds, GMCs), by 10$^4$ years, the $^{12}$C/$^{13}$C ratio is more constrained and  converging to a range between 40 and 70  with the exception of H$_2$CO whose \Cratio ratio is very low until past a million years when it then converges towards the canonical value (see Figure~\ref{fig:gal} Bottom).   Similarly, at densities of 10$^5$ cm$^{-3}$ and high visual extinction (simulating perhaps the densest gas within GMCs), the lowest $^{12}$C/$^{13}$C ($\sim$ 18) is found for  H$_2$CO (at least until 10$^6$ years), followed by HCN with a constant value of 40 (see Fig.~\ref{fig:denseGMC}). CS on the other hand has a variable and high \Cratio until 1 million years when it settles to the canonical value of 70.  In general, even for relatively quiescent environments, chemical fractionation can lead to a relatively large range of \Cratio across molecules in spirals. 
\par
\begin{figure}
   \includegraphics[scale=0.6]{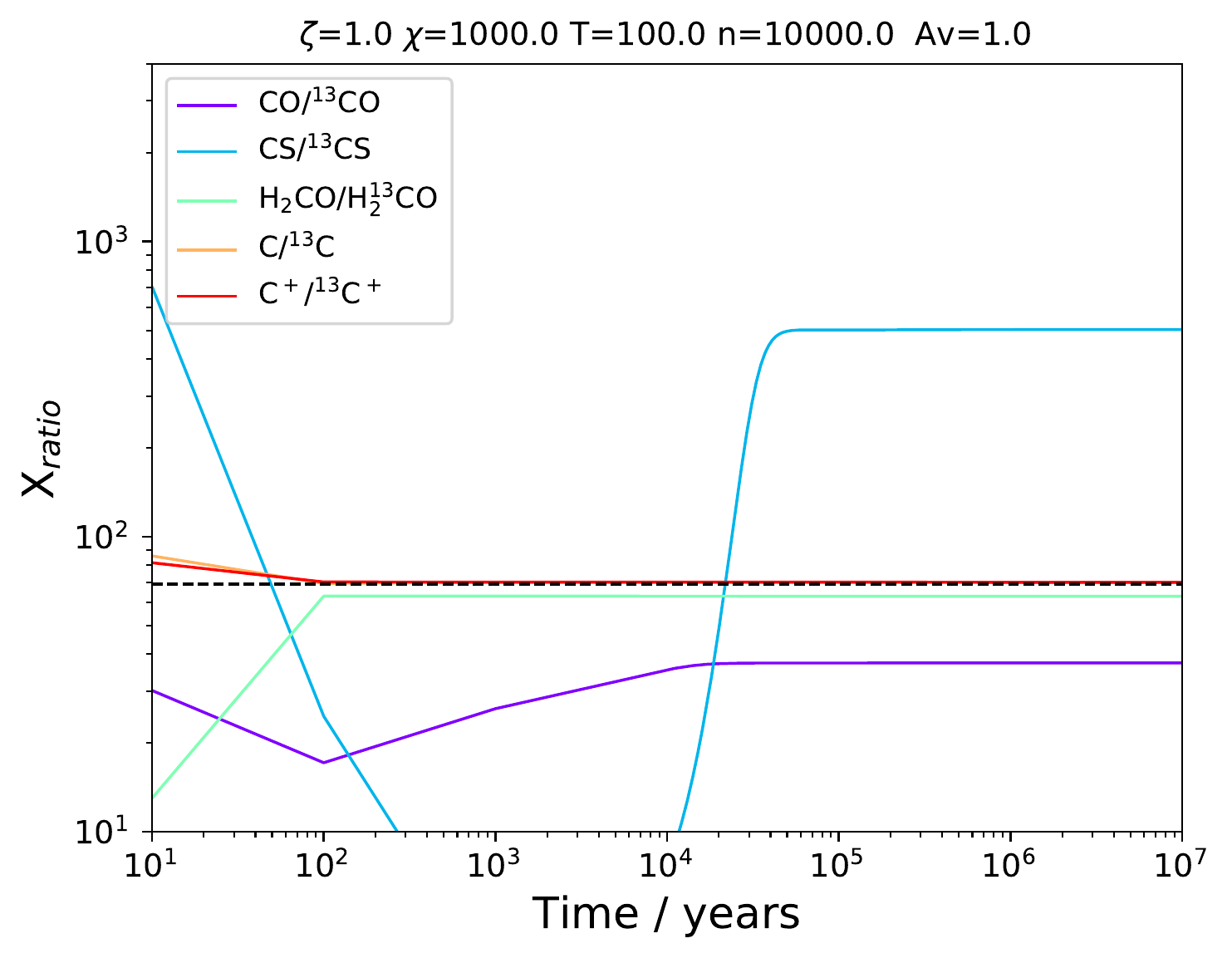}
   \includegraphics[scale=0.6]{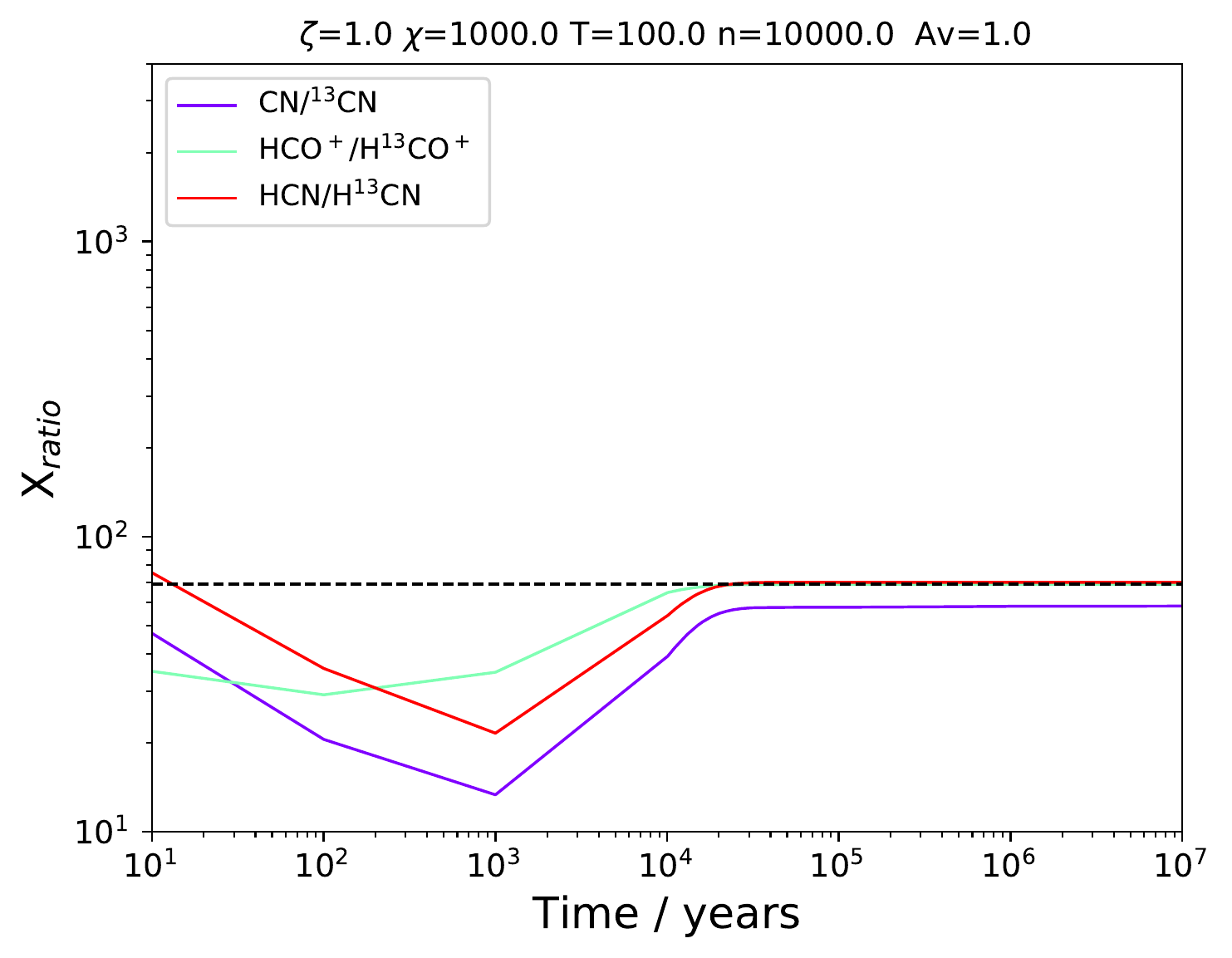}
\caption{\Cratio for selected species in extended gas in starburst galaxies.}
\label{fig:9str}
\end{figure}

{\it Starburst galaxies:}  in chemical models it is often accepted that one may reproduce such environments by using an FUV radiation
field enhanced as compared to normal spiral galaxies (Israel
\& Baas 2001; Bayet et al. 2008, 2009). Taking a $\chi$ = 1000 $\chi_o$ as representative we find that in gas at low visual extinction, FUV dissociation can highly influence fractionation across molecules,  including CO. For example, at 10$^4$ cm$^{-3}$ the $^{12}$C/$^{13}$C ratio is never higher than 70 for  all molecules but CS, where in the latter it  reaches values of 500 at steady state. The \Cratio ratio for CO is $\sim$ 40. (see Fig~\ref{fig:9str}). 

\begin{figure}
   \includegraphics[scale=0.6]{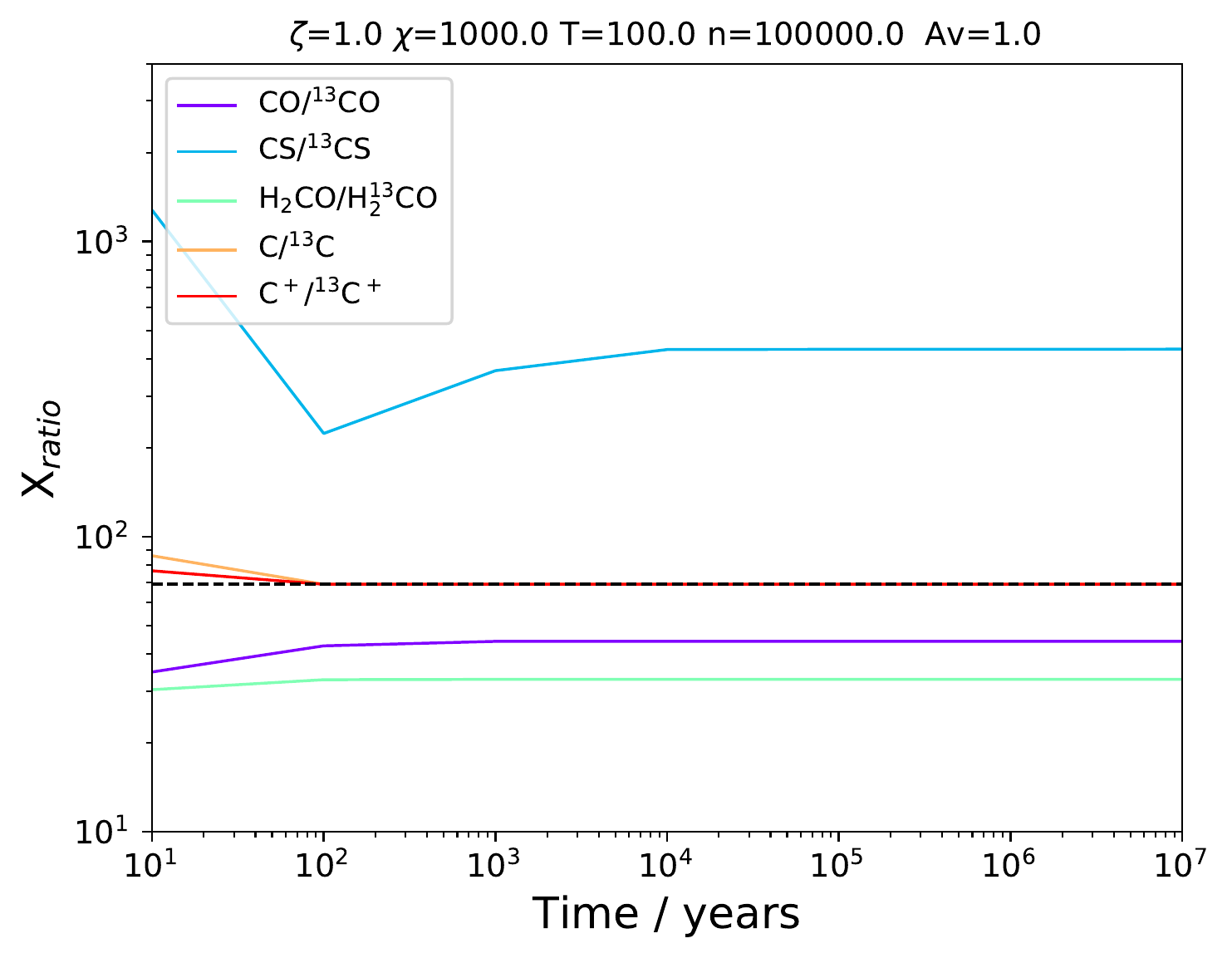}
   \includegraphics[scale=0.6]{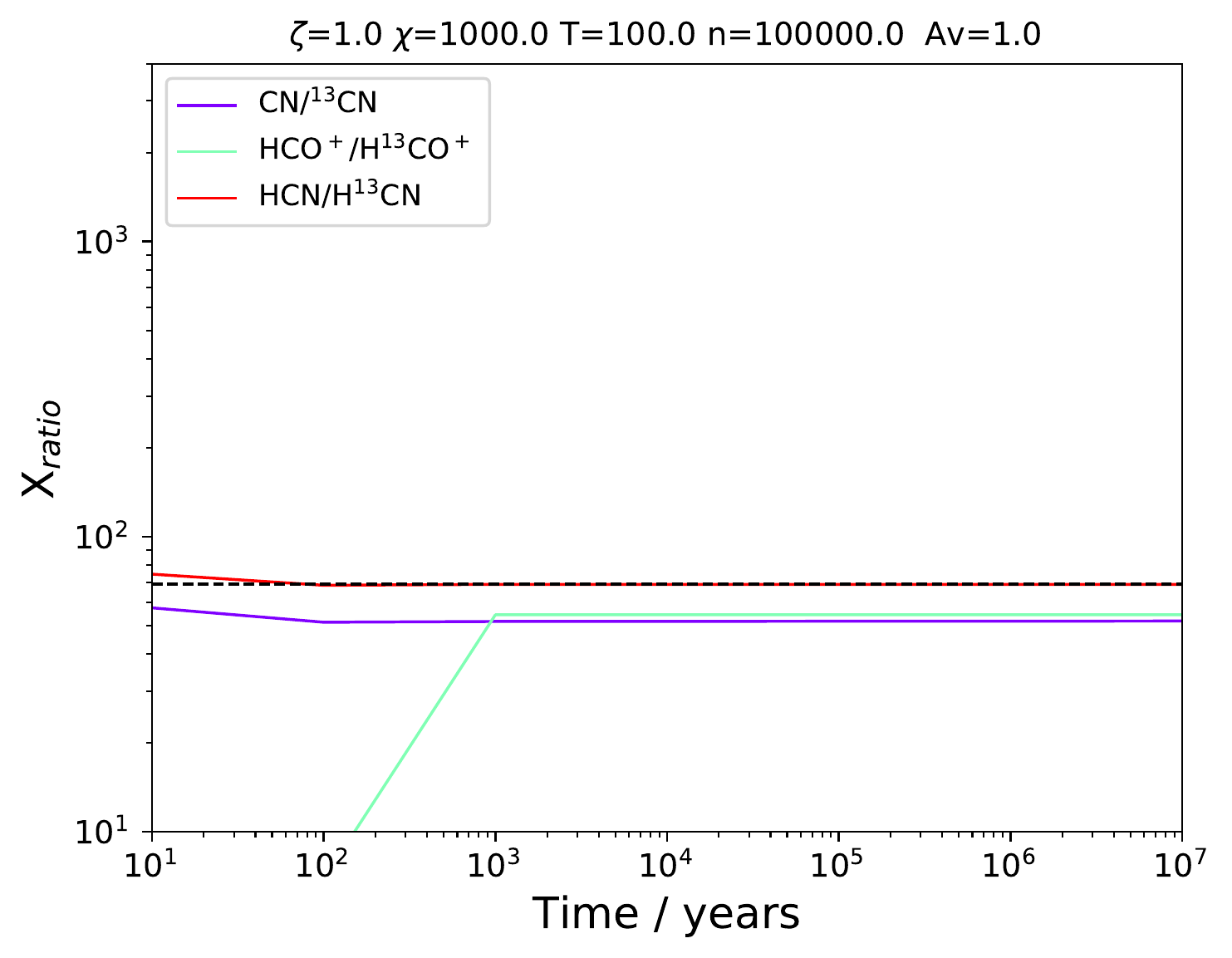}
\caption{\Cratio for selected species in dense  gas in starburst galaxies.}
\label{fig:21str}
\end{figure}

At higher densities, at low visual extinction,  the \Cratio ratio is much more constant with time. The differences between CS and the rest of the molecules are also quite large, with most species  having a $^{12}$C/$^{13}$C  below  70, and  the CS \Cratio ratio being as high as 400. The HCN molecule is the only one with a \Cratio ratio of 70. This is an interesting result as it implies that dense gas in starbursts should maintain a relatively constant, lower than solar, \Cratio,  for most species (see Fig~\ref{fig:21str}).

\begin{figure}
   \includegraphics[scale=0.55]{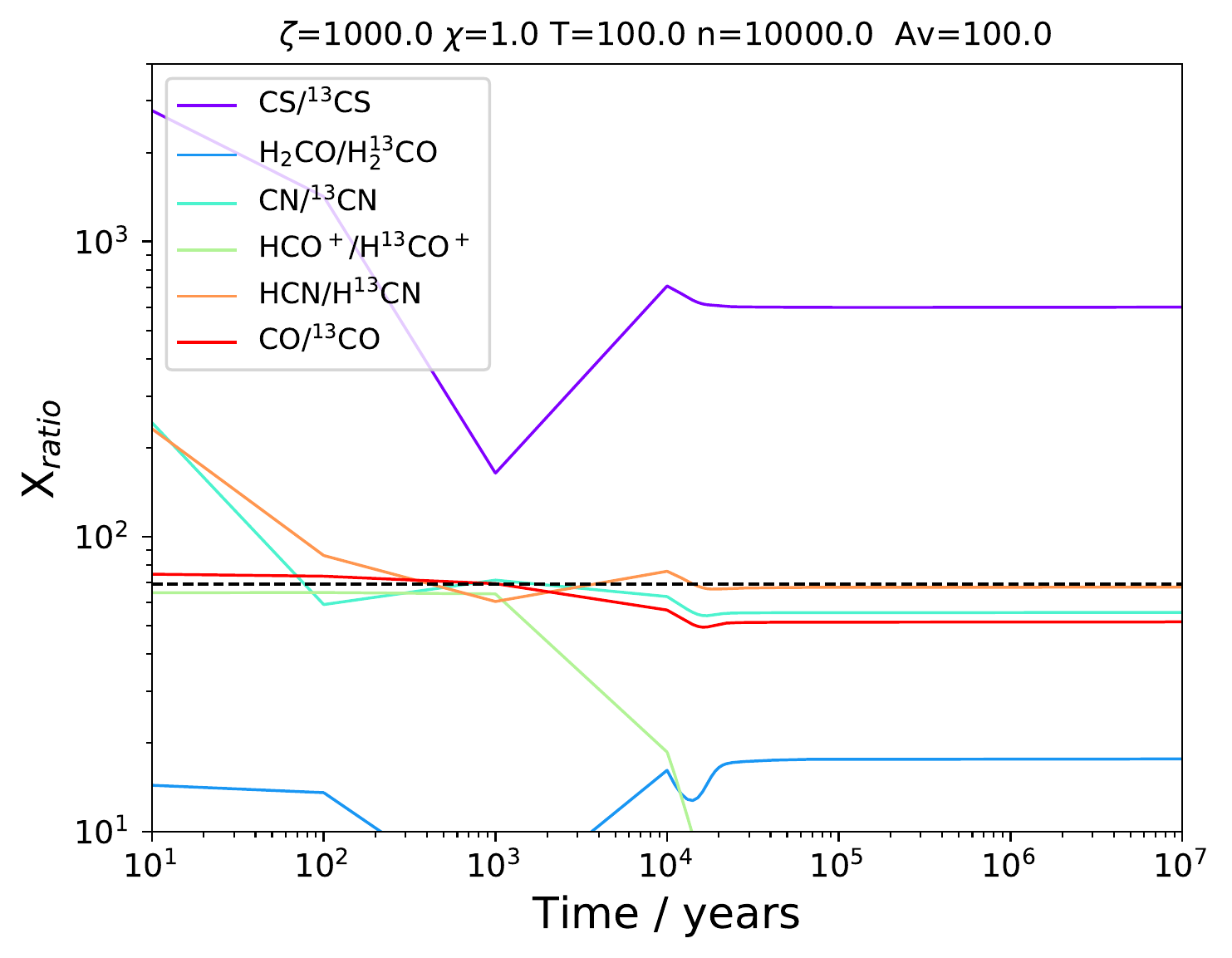}
   \includegraphics[scale=0.55]{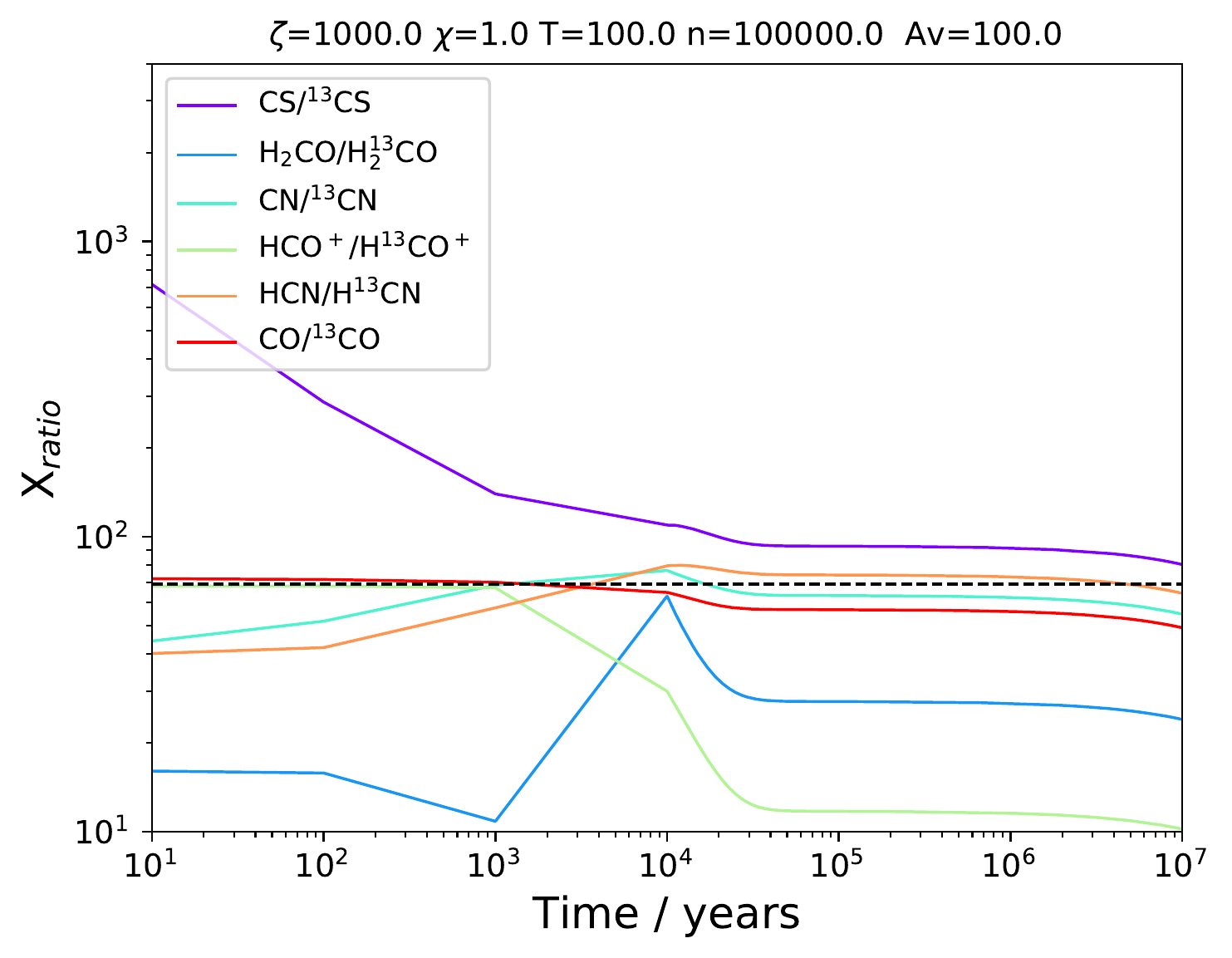}
\caption{\Cratio for selected species in AGN dominated galaxies. See also Figure~\ref{fig:Tvar}}
\label{fig:agn1}
\end{figure}

{\it AGN-dominated galaxies:} we choose to represent AGN dominated galaxies by models with a cosmic ray ionization rate 1000 times the Galactic one. As in Paper I, we do note that for the purpose of our modelling, the cosmic ray ionization flux parameter is also used to "simulate" an enhancement in X-ray flux. While this approximation has its limitations, the chemistry arising from these two fluxes should be similar (Viti et al. 2014).  Examples of such environments are then in Figure~\ref{fig:Tvar} at low A$_V$, and Figure~\ref{fig:agn1} at higher A$_V$. Interestingly, in these environments  the \Cratio ratio of CO is in fact always lower than 69. The  CS \Cratio ratio can instead reach  high values at densities of 10$^4$ cm$^{-3}$ only. We expect the \Cratio ratio for HCN and CN to be very similar in AGN-dominated galaxies,  while for H$_2$CO is always lower than for the rest of the species. 

\begin{figure}
   \includegraphics[scale=0.55]{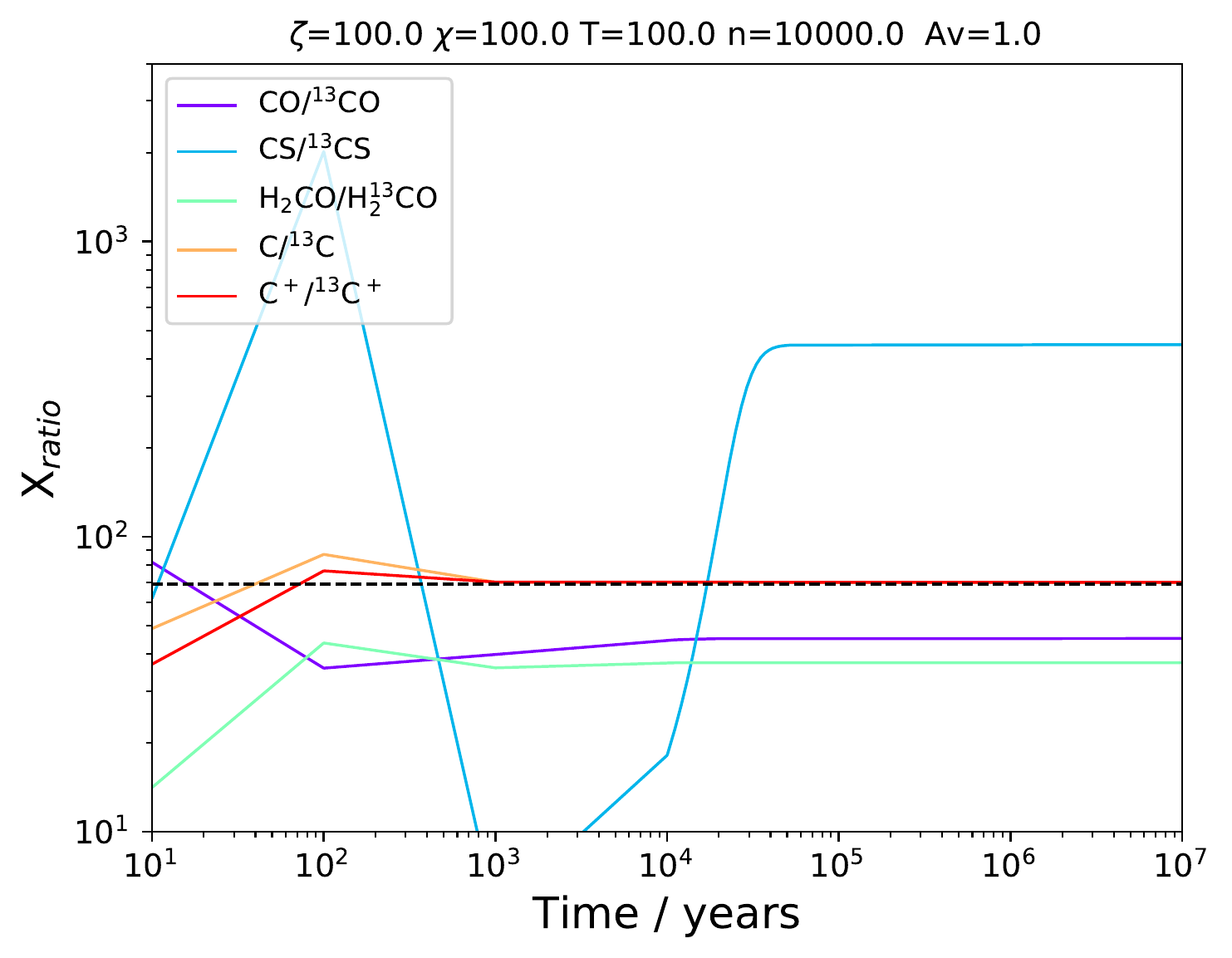}
   \includegraphics[scale=0.55]{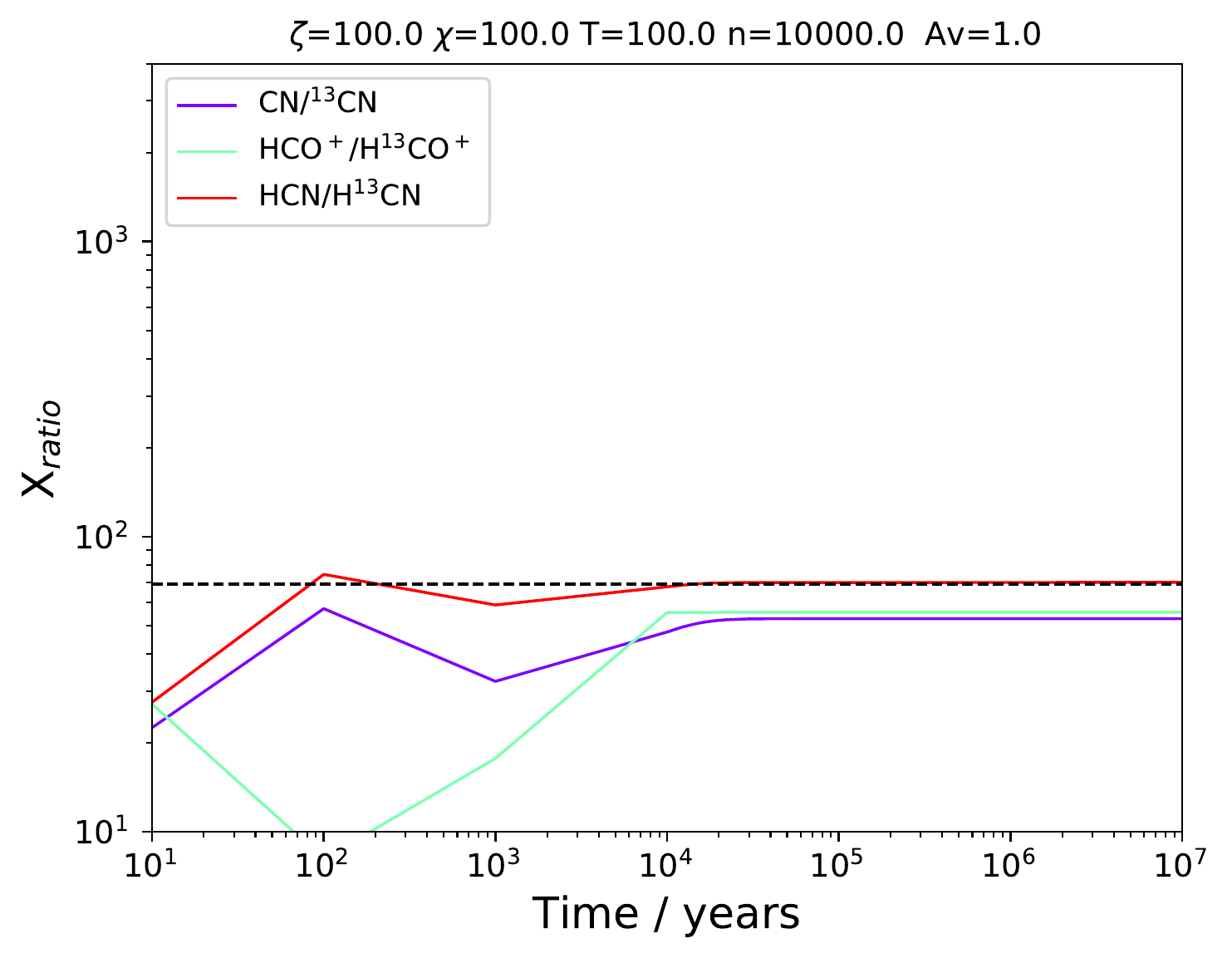}
\caption{\Cratio for selected species in composite galaxies.}
\label{fig:composite1}
\end{figure}

{\it Composite Galaxies:} These galaxies (where both an AGN and starbust rings are present) may be represented by models where both the cosmic ray ionization rate and the radiation field are enhanced. Obviously there are too many combinations to discuss here but if we were to take a representative combination of an enhancement of 100 for both $\zeta$ and $\chi$ then we find that at low  visual extinctions the $^{12}$C/$^{13}$C ratio is  again 69 only for HCN, with the CO and CN ones always lying below this value, at $\sim$ 40-50, and the CS one being as high as 400.  (see Fig.~\ref{fig:composite1}). 
At higher visual extinctions, the \Cratio for all species stabilizes   at $\sim$ 70 for most species as long as the gas density is above 10$^4$ cm$^{-3}$, while at lower densities it decreases for some species (H$_2$CO and HCO$^+$).

In addition to CO, our models suggest that HCN is the best molecular tracer  of the primordial \Cratio ratio, possibly instead of CN, as proposed by Henkel et al. (2014). This is because it suffers less from chemical fractionation than the other C-bearing species. While CO probes lower-density gas, HCN is therefore the best tracer to measure the \Cratio ratio toward high-density regions as in the nuclei of external galaxies.


\subsection{Comparison with observations}
\label{observations}

The aim of this section is to qualitatively determine whether there is any observational evidence that chemical fractionation can affect the $^{12}$C/$^{13}$C ratio in the various species in galaxies. Of course we  note that  the elemental isotope ratio can be very sensitive to IMF variations (Romano et al. 2017, 2019) which would skew it from its canonical value of $\sim$ 70 before any chemistry can take place in the ISM.

In Table 2, we list some of the observational values of \Cratio\ for  external galaxies reported in the literature. In addition to this table we also note a study of carbon-bearing species across a large sample of ULIRG and starburst galaxies performed by Israel et al. (2015) who find a very high $^{12}$CO/$^{13}$CO ratio (200--3000). 

In this context, we briefly discuss our findings below, bearing in mind that most of the measurements obtained so far encompass the whole size of the galaxy, hence any chemical fractionation will be the consequence of global and not local effects. Even in the very few studies performed so far at high-angular resolution, e.g. the recent ALMA maps of Mart\'in et al. (2019) and Tang et al. (2019), the spatial scales resolved are of the order of $\sim 100$~pc, which may include star-forming as well as quiescent gas, and hence caution needs to be taken when drawing conclusions on the origin of the chemical fractionation. Moreover, as a further caveat  we note that the elemental isotopic ratio, in starbursts and high redshift galaxies in particular, may significantly differ from the one used here. Our comparison is, on the other hand, always made assuming that the elemental ratio at the beginning of the calculation is not too different from the galactic one. 

We further note that while, ideally,  every isotopologue comparison below should be accompanied with one from the main species, this is seldom possible as the main species is either not reported  or often derived from different observations. More importantly, direct comparison of single species between models and observations can not be performed as the beam, even at the highest spatial resolution, will encompass multiple gas components (this is discussed at length in e.g. Viti 2017). Comparisons of the \Cratio ratio on the other hand removes the uncertainties of the beam dilution, provided one assumes the two species arise from the same gas. Nevertheless, for transparency, for one of the models discussed below, we plot the abundances of the main species and show that they are indeed reasonable in their range of values, and  indeed qualitatively comparable to what one would expect or find for that type of galaxy.


\begin{itemize}
    \item {\it Starburst Galaxies}: 
  For this type of galaxies we compare the \Cratio for CN and CS with our models. For CN we in fact find that, at steady state, all models predict a \Cratio that falls in the range of 30 to 67 and hence this comparison is not informative. The carbon fractionation of CS  may be  uncertain due, among other things, to the uncertainties in the fractionation of sulfur (Henkel et al. 2104; Chin et al. 1996). Values have been reported ranging from $\sim$ 27 to $\sim$ 70.  We find that this range is satisfied by many models from our grid, and they all have a radiation field larger than 1 Draine. Moreover if we exclude models whose \Cratio ratio for CS only falls into the required range after at least 10$^4$ years, then the remaining models all have a cosmic ray ionization rate higher than Galactic and a gas density of at least 10$^5$ cm$^{-3}$. This combination of parameters is consistent with that expected for starburst galaxies and for gas traced by the high density tracers such as CS. We show in Figure~\ref{fig:CSstar} the abundance of the main isotopologue of CS, as well as other standard abundant molecular species, for one of the models whose \Cratio fit the observations.

\begin{figure}
\hspace{-0.8cm}
   \includegraphics[scale=0.6]{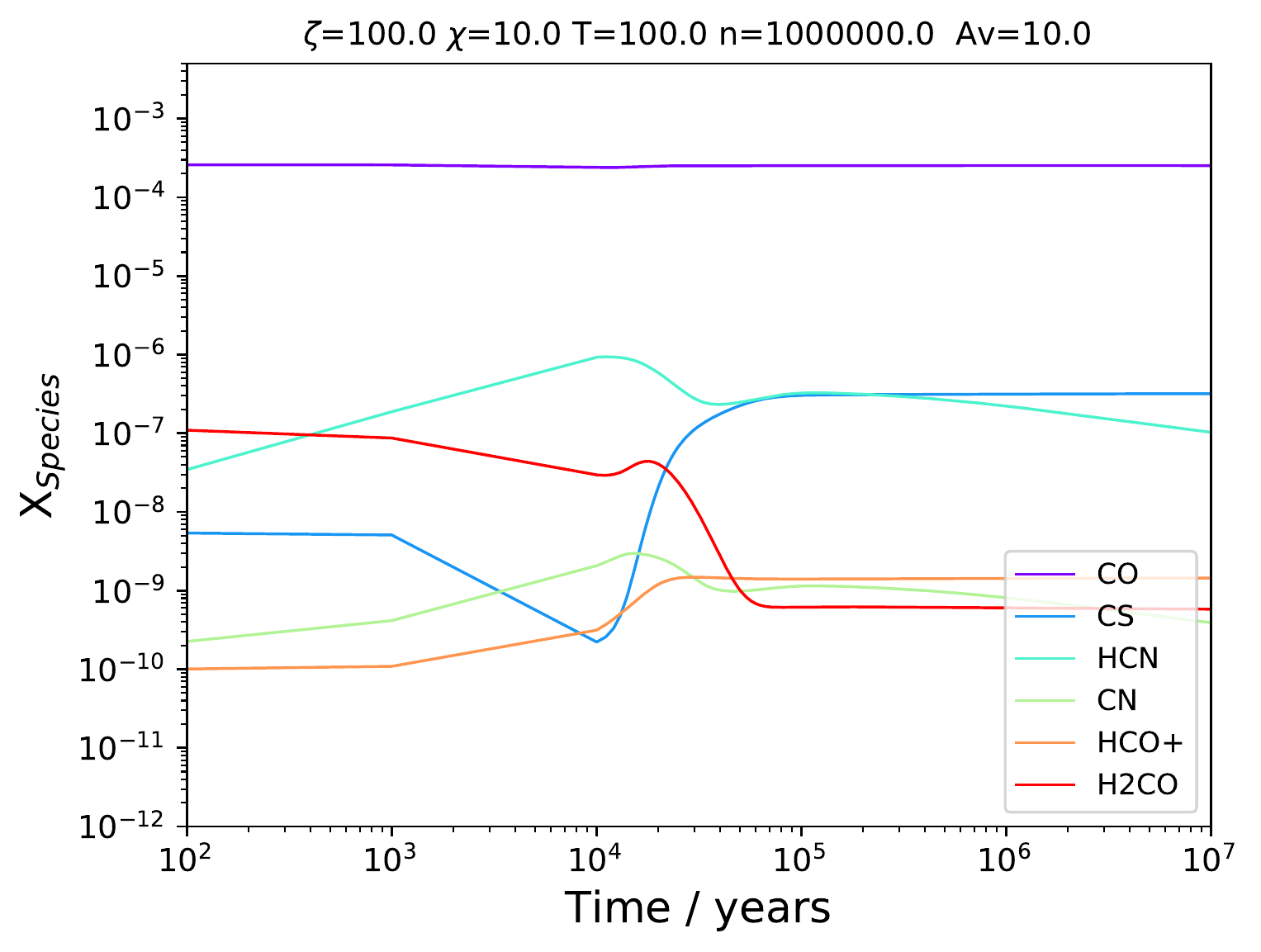}
\caption{One of the best matching models for the CS fractionation in starburst galaxies. }
\label{fig:CSstar}
\end{figure}

\item{\it Local and high redshift LIRG/ULIRG}: for these galaxies, CO is the only species for which we have a  \Cratio, except for Mrk 231 for which the \Cratio ratio has been derived also from CN, and it is $\sim$ 100. In all cases  the CO \Cratio ratio is measured to be in the range $\sim$ 100--230. This range is never matched although for the first 100 years or so there are some low density and low visual extinction models that approach a \Cratio ratio of 100.  The CN \Cratio ratio is also never matched by our models at steady state, although models with a high visual extinction and a high density match this ratio at early times ($<$ 10,000 years).  We believe that the failure of chemical models to match the high \Cratio ratios for CN and CO is a confirmation that the isotopic decrease is  $not$ a product of ISM chemistry but rather of nucleosynthesis..


\item{\it LMC:} for this low-metallicity galaxy the \Cratio is derived from H$_2$CO and it is $\sim$ 50.  Since the metallicity for this galaxy is believed to be half of the solar one, we have ran a full grid of models at such metallicity. Several models perfectly match this ratio. The common parameters to $all$ matching models are a high cosmic ray ionization rate  ($>$ 10 $\zeta_o$) and a low visual extinction (1 mag). See Fig.~\ref{fig:lmc} for examples of best fit models. 
\begin{figure}

   \includegraphics[scale=0.5]{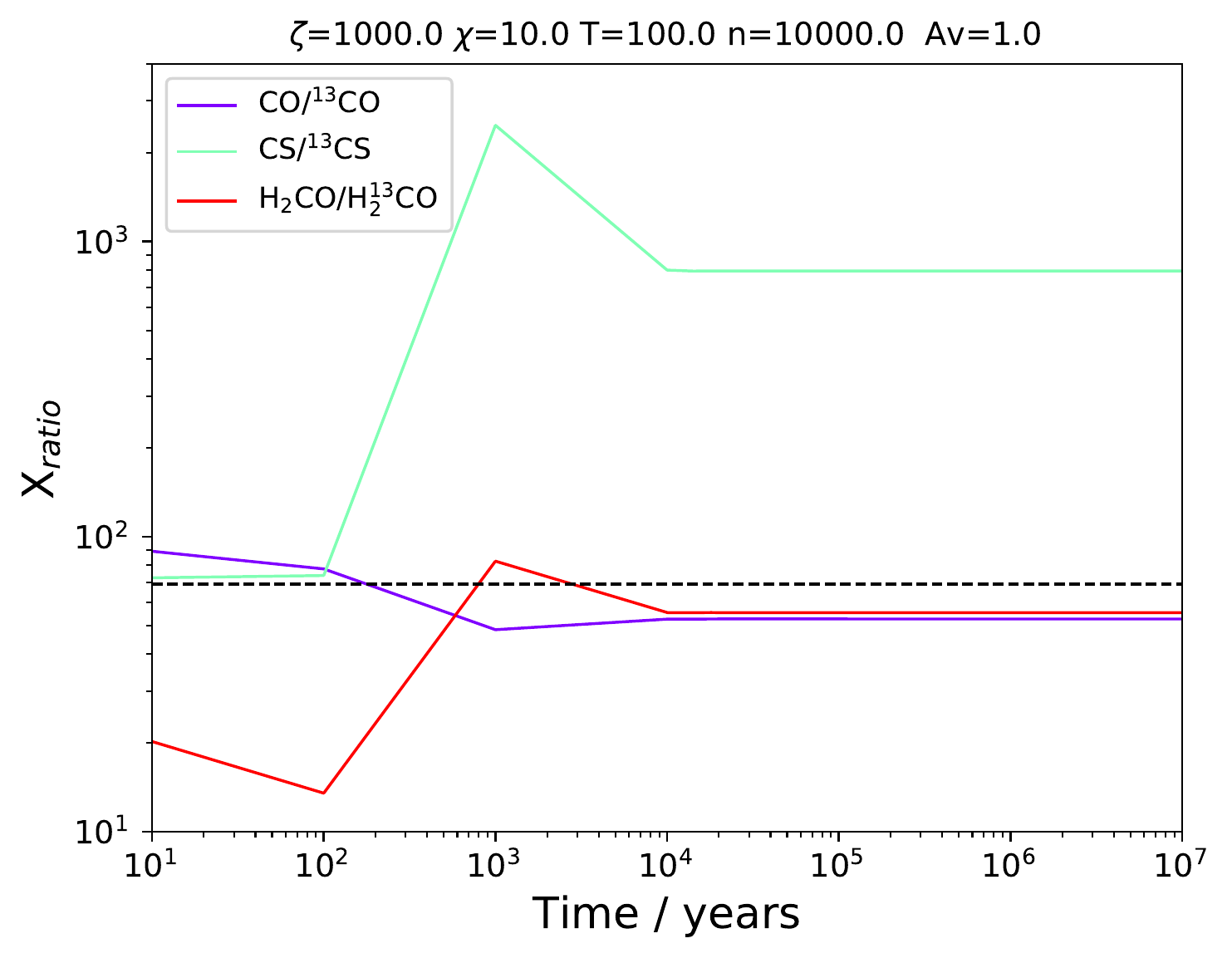}
   \includegraphics[scale=0.5]{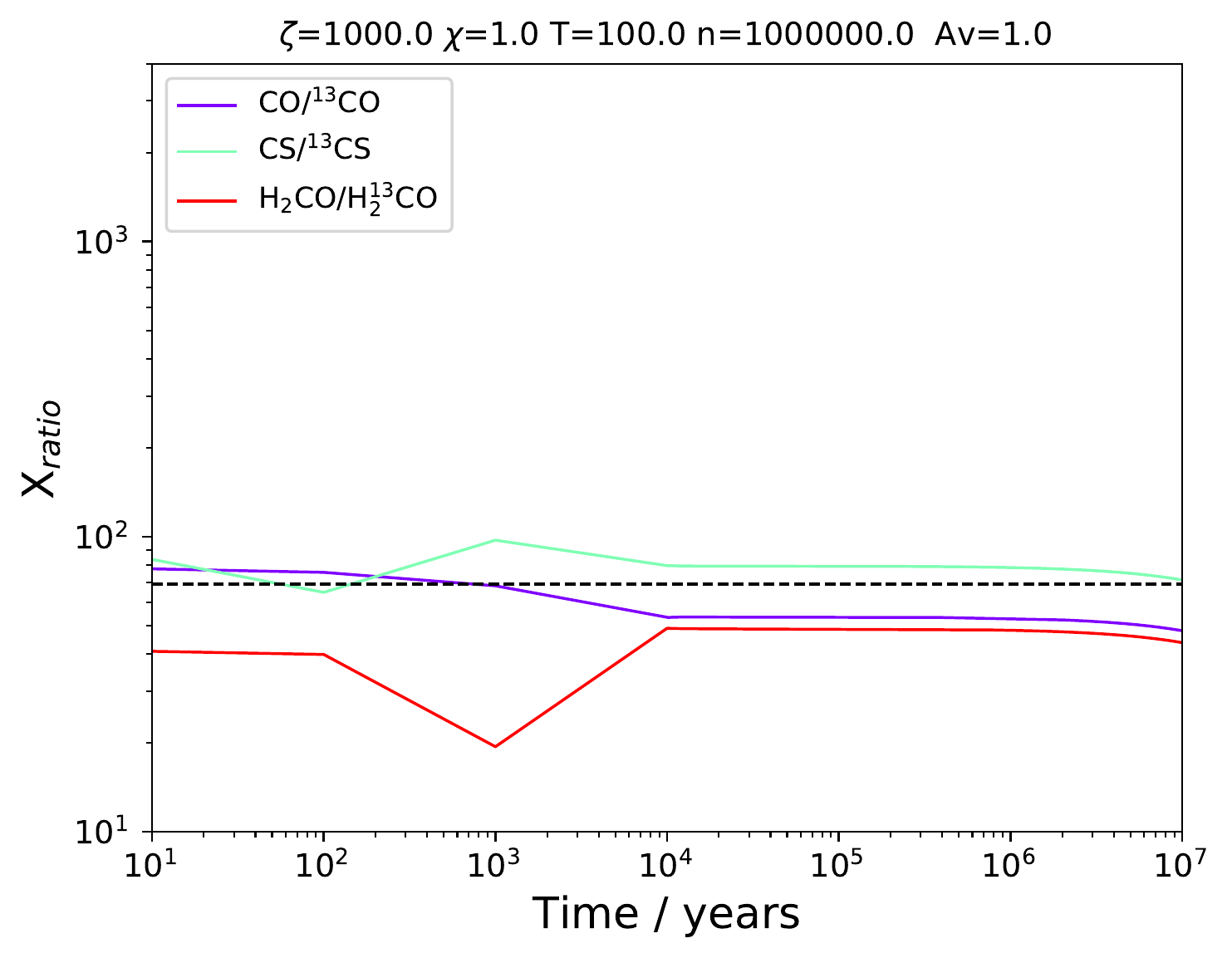}
   \caption{Two models at half the solar metallicity,  where the \Cratio for H$_2$CO approaches $\sim$ 50.}
   \label{fig:lmc}
\end{figure}

\item{\it AGN/composite galaxies:} for the composite galaxy NGC~1068 our only measure of carbon fractionation comes from CN. A \Cratio ratio of $\sim$ 50  can be matched by many models, $all$ sharing two common parameters: a visual extinction of 1 mag and a radiation field higher than 1 Draine. This finding is consistent with CN being a tracer of PDR gas.  Aladro et al. (2013)  reports values for the column density of the main isotopologue of CN of 5-6$\times$10$^{15}$ cm$^{-2}$ which, at a visual extinction of 1, would correspond to a fractional abundance of 3$\times$10$^{-6}$. It is not possible to reach such a high fractional abundance of CN by a single PDR. It is much more likely that within the beam of the Aladro et al. (2013) observations a number of PDRs were present: of the models that match a \Cratio ratio of 50 for CN, several can reach up to 5$\times$10$^{-8}$--10$^{-7}$ in the CN fractional abundance which would imply $\sim$ 10-50 PDRs within the beam size. For the other three active galaxies from Table 2 all give a \Cratio ratio of $\sim$ 40 to $\sim$ 45 for HCO$^+$ and HCN, but higher by almost a factor of 100 for CO (for at least one galaxy). A ratio of $\sim$ 40 for HCO$^+$ is hard to match although a model with a very high cosmic ray ionization rate, a density of 10$^5$ cm$^{-3}$ and again a visual exctintion of 1 mag would be the best match. On the other hand a ratio closer to $\sim$ 50 for HCO$^+$ is  reached as long as  the visual extinction is again 1 mag, and the radiation field higher than Galactic, consistent with our findings from CN. Figure~\ref{fig:CN-HCOcompo} shows the \Cratio ratio for HCO$^+$ and CN  for one of our best fit models. On the other hand, a \Cratio ratio of 40 for HCN is only achieved for models  with a high density of 10$^6$ cm$^{-3}$ and a high visual extinction (see Figure~\ref{fig:denseGMC}), which could indicate $^{13}$C enrichment due to fractionation in the densest and not illuminated regions. As before, a \Cratio ratio of 100 for CO is never reached, excluding ISM chemistry as the cause of such a high ratio. We note in fact that Papadopoulos et al. (2014) finds a $^{12}$CO/$^{13}$CO ratio which is even higher (200-500) but they attribute this to the possibility that this is due to a top-heavy IMF sustained over long timescales in galaxies which are active (mergers or starburst). 


\begin{figure}
\hspace{-1.0cm}
   \includegraphics[scale=0.6]{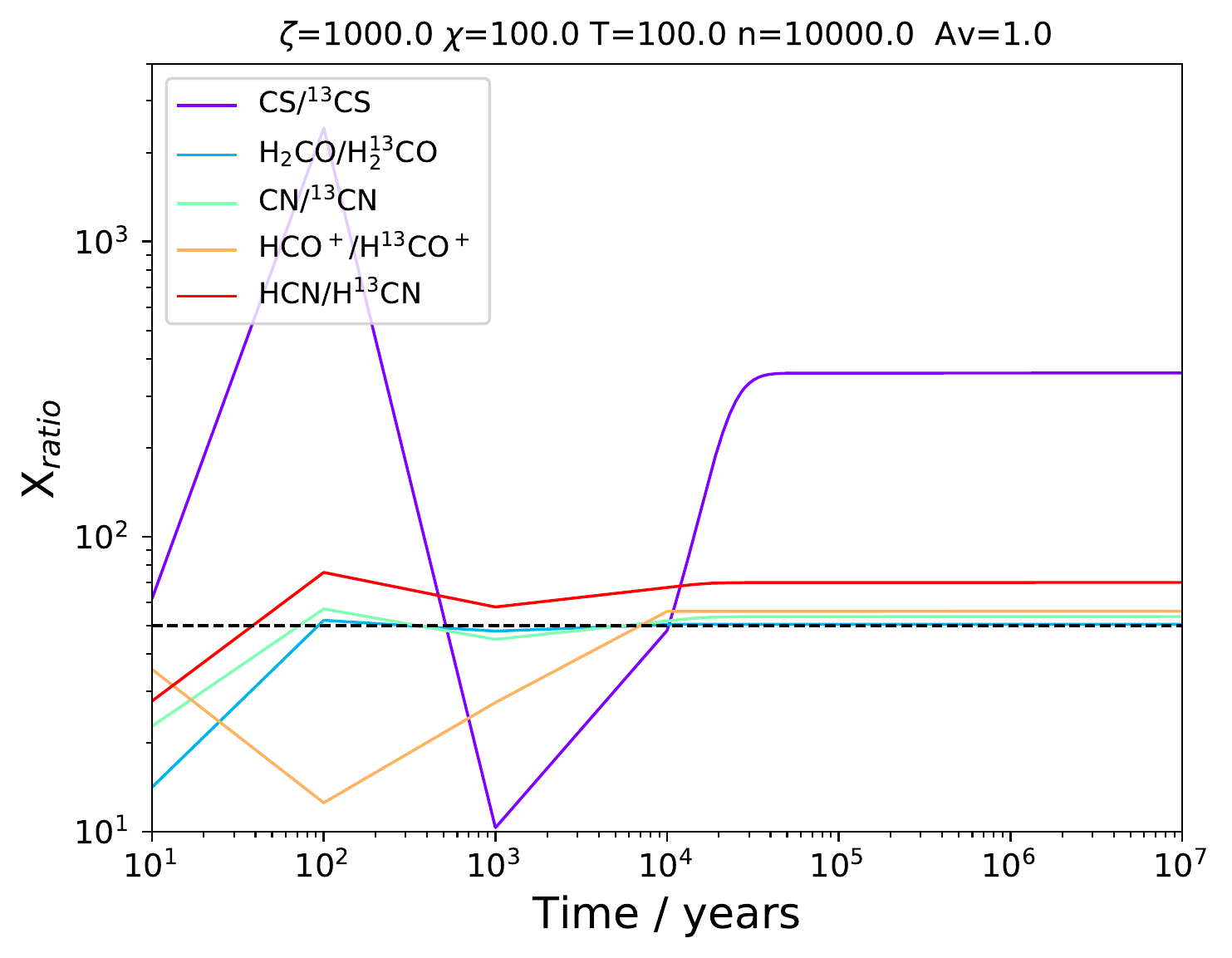}
   \caption{The CN and HCO$^+$ fractionation in composite galaxies for one of our best model fits.}
   \label{fig:CN-HCOcompo}
\end{figure}

\item{\it Spirals}: for the local spiral galaxy IC~342 we only have lower limits for the \Cratio ratio as derived from observations of CN. Our conclusions for NGC~253 are also valid for this case.  For the high redshift galaxies MA0.89 and MA0.68 the \Cratio was averaged from observations of HCN, HCO$^+$ and HNC from Muller et al. (2006) and Wallstr\"{o}m et al. (2016) to be $\sim$27 and $\sim$40, respectively.  There are no models that give either value for all of these species at the same time. In fact, especially for HCO$^+$, it is either lower than 27 or higher than 40. Without more knowledge of the individual values for each of these species for each galaxy we can not speculate any further, other than saying that the models that produce a similar \Cratio for all the three species at $\sim$ 50 all have a cosmic ray ionization rate of 1000 $\zeta_o$ and a gas density of 10$^6$ cm$^{-3}$.  As the Muller et al. (2006) and Wallstr\"{o}m et al. (2016) observations are absorption measurements of the intervening gas in  spiral galaxies along the line of sight of quasars, it is very unlikely that the detected molecules are tracing such a high density gas.  

\end{itemize}

\section{Conclusions}
In this paper we report the results of a theoretical chemical study of the carbon fractionation in the ISM, where we investigated the effects of chemical fractionation under different physical conditions, resembling the range of conditions found in external galaxies. Our main conclusions are:
\begin{enumerate}
    \item As for the case of nitrogen fractionation (see Paper I), a difference of a factor of 2 in gas temperature (from 50 to 100 K) does not seem to affect chemical fractionation much (a factor 2 in the most extreme cases).
    \item  Under the conditions investigated, the $^{12}$CO/$^{13}$CO ratio is relatively constant, and in most cases it is close to 70. 
    \item  For species other than CO, carbon fractionation has a strong dependence on the time evolution of the gas as well as on the physical parameters employed, although HCN seems to be the best tracer to probe the \Cratio ratio in the majority of cases. On the other hand we note that H$_2$CO often exhibits very low fractionation, and may therefore not be a good species for measuring the \Cratio ratio. 
    \item A high cosmic ray ionization rate and/or a high radiation field leads to a much more variable fractionation  for most species.
    \item As previously found by other authors (e.g. Roueff et al. 2015) there are always large variations in the \Cratio among the carbon bearing species investigated.
    \item Qualitatively, we are able to match most of the observations of the \Cratio in external galaxies. We note however that our models do not take into consideration differences in nucleosyntesis products i.e we assume that the elemental isotopic carbon ratio is the so called standard one (Wilson 1999). 
    \item Our model predictions, when compared to observations, strongly support the theory that a high \Cratio in CO ($>$ 100) is a product of nucleosynthesis rather than ISM chemistry. 
    \item As all the observations were performed with a beam that encompasses very extended gas, any variation in fractionation is not due to local conditions (e.g star formation versus quiescent gas) but to the global effects of chemical fractionation. 
\end{enumerate} 

We stress once more that the few current measurements cannot constrain linear scales smaller than $\sim 100$~pc, and hence only very high-angular resolution observations will allow us to observe local, instead of global, isotopic ratios and put more stringent constraints on our models.

\begin{table*}
\begin{center}
\caption{\Cratio measured in external galaxies}
\begin{tabular}{lcccc}
\hline \hline
Galaxy  & type & $^{12}$C/$^{13}$C  &  Molecule & Reference  \\           
\hline
NGC 253 & starburst & $\sim$ 40; $30-67$ & CN & Henkel et al. (2014); Tang et al. 2019 \\
NGC 253 & starburst & $\sim 27-70$ & CS & Martin et al. (2005, 2006); Henkel et al. (2014) \\
M82    & starburst & $>40$ & CN & Henkel et al. (1998) \\
NGC 253 & starburst nucleus & $\sim 21$ & C$^{18}$O$^a$ & Mart\'in et al. (2019) \\
NGC 4945 & starburst nucleus & 6--44 & CN & Tang et al. (2019) \\
\hline
         & starburst & $\sim 10 - 70$ & & \\
\hline
VV 114   & LIRG & $\sim 230$ & CO & Sliwa et al. (2013) \\
NGC 1614 & LIRG & $\sim 130$ & CO & Sliwa et al. (2014) \\
Mrk 231 & ULIRG    & $\sim 100$ & CO, CN & Henkel et al. (2014) \\
Arp 220 & ULIRG   & $\sim 100$ & CO & Gonz\'alez-Alfonso et al. (2012) \\
Arp 193 & ULIRG   & $\sim 150$ & CO & Papadopoulos et al. (2014) \\
Cloverleaf & ULIRG $z=2.5$ & $100-200$ & CO & Spilker et al. (2014) \\
Eyelash    & ULIRG $z=2.3$ & $\sim 100$ & CO & Danielson et al. (2013) \\
\hline
   & LIRG/ULIRG  & $\sim 100 - 230$ & & \\
\hline
LMC       &  0.5 metal & $\sim 49$ & H$_2$CO & Wang et al. (2009) \\
\hline
NGC 1068 &  AGN+starburst  & $\sim$ 50 ;24--62 & CN & Aladro et al. (2013); Tang et al. (2019) \\
NGC 4258 & AGN & $\sim 46$ & HCO$^+$ & Jiang et al. (2011) \\
NGC 3690 & AGN+starburst & $\sim 40$ & HCO$^+$ & Jiang et al. (2011) \\
NGC 6240 & AGN+starburst & $\sim 41$ &  HCN & Jiang et al. (2011) \\
NGC 6240 & AGN+starburst & $300-500$ &  CO &  Papadopoulos et al. (2014); Pasquali et al. (2004) \\
\hline
    & AGN/composite & $\sim 20-60$ (except NGC 6240 in CO) & & \\
\hline
IC 342 & spiral local & $>30$ & CN & Henkel et al. (1998) \\
MA0.89 & spiral $z=0.89$ & $\sim 27$ & HCN, HCO$^+$, HNC & Muller et al. (2006) \\
MA0.68 & spiral $z=0.68$ & $\sim 40$ & HCN, HCO$^+$, HNC & Wallstr\"{o}m et al. (2016) \\
\hline
       & spiral & $\sim 30-40$ & & \\
\hline
\end{tabular}
$^a$ In this case the \Cratio ratio is extracted from the $^{12}$C$^{18}$O/$^{13}$C$^{18}$O \\
\end{center}
\end{table*}

\section*{Acknowledgements}

SV  acknowledges support from the European Research
Council (ERC) grant MOPPEX ERC-833460. SV and FF acknowledge the support of STFC (grant number ST/R000476/1) and the Center for Astrochemical Study at MPE for partially supporting this work. I.J.-S. has received partial support from the Spanish FEDER (project number ESP2017-86582-C4-1-R). We thank Prof Roueff for providing us the needed information to perform a benchmarking with her model,  and the anonymous referee for their constructive comments which improved the manuscript.  

\section*{Data availability}
The model outputs underlying this article will be freely shared on request to the corresponding author.




\bsp	
\label{lastpage}
\end{document}